# Spin-Polarized Oxygen Evolution in Chiral-Molecule-Modified Plasmonic Photoanodes

Priscila Vensaus[1], Milad Sabzehparvar[1], Fatemeh Kiani[1], Germán García Martínez[1], Giulia Tagliabue[1*]

1) Laboratory of Nanoscience for Energy Technologies (LNET), STI, École Polytechnique Fédérale de Lausanne, 1015 Lausanne, Switzerland

*E-mail: giulia.tagliabue@epfl.ch

## Abstract

Photoelectrochemical oxygen evolution is limited not only by multi-electron charge-transfer kinetics but also by the spin constraints associated with forming triplet $O_2$. Here, we used a hybrid photoanode architecture to demonstrate spin-polarization of plasmonic hot holes through a chiral molecular layer. $TiO_2$ photoanodes were modified with achiral Au nanoparticles to introduce visible-light plasmonic absorption, functionalized with cysteine as a chiral molecular interface, and coated with a NiFe-based oxygen-evolution catalyst. Wavelength-resolved photo-scanning electrochemical microscopy was used to directly detect locally evolved $O_2$ under operando illumination while simultaneously monitoring the photoanode current. Chiral functionalization with homochiral L-cysteine enhanced both photocurrent and local $O_2$ evolution relative to racemic DL-cysteine controls. The chirality-dependent enhancement was most pronounced under visible excitation overlapping the Au plasmon resonance, including a 130% photocurrent increase. These results provide evidence that chiral molecular layers, often used for chiral nanoparticle synthesis, can directly modulate plasmon-derived hot-carrier transfer through the chiral induced spin selectivity (CISS) effect. This work establishes a chiral plasmonic photoelectrochemical platform for coupling hot-carrier generation to spin-dependent water oxidation.

## Introduction

Solar-driven conversion of small molecules, such as water and $CO_2$, into fuels and value-added chemicals is a central goal in the development of sustainable energy technologies.[1] Photoelectrochemical (PEC) solar fuel production is particularly attractive because it offers a direct route to couple light absorption, charge separation, and interfacial redox chemistry within a single device architecture. Among the relevant half-reactions, the oxygen evolution reaction (OER) remains one of the most demanding because it requires the concerted removal of four electrons and four protons from water to form $O_2$, leading to substantial overpotentials.[2,3] In photoelectrochemical systems, these intrinsic catalytic limitations are further coupled to the need for efficient photon absorption, charge-carrier generation, transport, and transfer across the semiconductor/electrolyte interface.

An increasingly important aspect of OER is its spin dependence.[4] The change in spin multiplicity when converting water (singlet) to oxygen (triplet), implies that OER is not governed solely by charge-transfer energetics and binding energies, but may also be influenced by the spin character of the charge carriers and reaction intermediates involved in O–O bond formation. Recent studies have therefore highlighted spin polarization, magnetic interactions, and spin-selective charge transfer as emerging design parameters for water oxidation catalysts.[4–6]

Chiral molecular interfaces represent a promising route for inducing spin-selective charge transfer through the chirality-induced spin selectivity (CISS) effect.[6,7] When a chiral molecular layer is coupled to an OER catalyst, spin-selective charge transfer can favour reaction pathways that facilitate the formation of triplet oxygen, thereby reducing spin-related limitations.[8,9] Indeed, several recent dark electrochemical studies have shown that chiral molecule functionalization can strongly enhance OER activity on catalytic electrodes, including Ni-, NiFe- and FeCo-based systems.[8–10] However, demonstrations of this concept in photoelectrochemistry remain sparse.[11,12]

In PEC OER, $TiO_2$ remains a benchmark photoanode material because of its abundance, chemical stability, low toxicity and suitable band-edge positions.[13] However, its large band gap restricts absorption primarily to the UV region, limiting its ability to harvest the solar spectrum efficiently. $TiO_2$ sensitization using metallic nanoparticles that support localized surface plasmon resonances (LSPR) has emerged over the years as a viable strategy to broaden light harnessing to the visible/near infrared spectral range.[14–16] Plasmonic nanoparticles can in fact generate both strong local electromagnetic-field enhancements and energetic hot-carriers, promoting charge generation above and below the $TiO_2$ bandgap, respectively. For water oxidation, Au/$TiO_2$ systems have previously been shown to support plasmon-assisted OER under visible excitation, demonstrating that plasmonic excitation can be coupled to oxidative chemistry at semiconductor interfaces.[16–18]

Recent studies have explored the use of chiral plasmonic nanoparticles and reported chirality-induced enhancements in the OER, mainly in dark electrochemical studies.[19–21] However, disentangling the respective roles of nanoparticle chirality and the presence of chiral ligands used during synthesis remains challenging. Integrating chiral molecules with otherwise achiral plasmonic Au/$TiO_2$ photoanodes offers a more controlled strategy for coupling visible-light-driven hot-carrier generation with spin-polarized interfacial charge transfer, isolating the contribution of molecular chirality on plasmon-assisted photoelectrochemical OER independently of nanoparticle chirality. To the best of our knowledge, this architecture has not yet been reported.

In this work, we develop a hybrid photoanode that integrates $TiO_2$ as the stable semiconductor scaffold, Au nanoparticles as the achiral plasmonic sensitizers, NiFe as the catalyst for water oxidation,[22] and a chiral molecular functionalization as a spin-polarizing element. Using wavelength-resolved scanning photo-electrochemical microscopy (photo-SECM) as an operando approach to detect locally generated $O_2$ near the illuminated photoanode surface,[23] we show that homochiral L-cysteine-functionalized photoanodes generate higher photocurrents, evolve more oxygen and have higher internal quantum efficiency than racemic DL-cysteine controls. This response suggests that chiral molecular functionalization can spin-polarize plasmonic hot-carriers through the CISS effect during photoelectrochemical water oxidation, in structurally achiral plasmonic nanoparticles. By integrating wavelength-resolved oxygen detection with chiral molecular engineering, this work provides evidence for coupling plasmonic hot-carrier generation to spin-selective catalysis in photoelectrochemical water oxidation.

## 2. Results and discussion

### 2.1. Fabrication and characterization of photoanodes

40 nm thick $TiO_2$ on ITO/fused silica substrates were modified with Au nanoparticles followed by deposition of a NiFe-based catalyst layer (**Figure 1a**). The Au nanoparticles were obtained by dewetting from a sputtered 2 nm layer, and NiFe (Ni:Fe 81:19 wt.%) was evaporated with a nominal loading of 0.5 nm. Further details of the sample fabrication process are present in the **Methods section**.

SEM images confirmed the formation of dispersed Au nanoparticles on the $TiO_2$ surface, with a size distribution of 9 ± 3 nm (**Figure 1b and Figure S1**). SEM images after NiFe deposition show no obvious aggregation or large-scale morphological change of the Au nanoparticle distribution (**Figure S2**). Although NiFe is deposited as a nominal metallic layer, exposure to alkaline electrolyte and anodic operation are expected to oxidize the surface, likely generating Ni/Fe oxyhydroxide-like species capable of mediating OER, as seen in similar systems.[24] This behavior is supported by dark cyclic voltammetry of $TiO_2$/NiFe, which shows a distinct $Ni^{2+}/Ni^{3+}$ redox feature before the onset of significant OER current (**Figure S3**).[25,26] Therefore, throughout this work we refer to the deposited film as a NiFe-based catalyst layer, recognizing that the active surface under operating conditions is likely an oxidized phase rather than metallic.

Optical absorptance spectra measured in water (**Figure 1c**) show the contributions of Au nanoparticles and the NiFe catalyst to the light-harvesting properties of the $TiO_2$ photoanodes. $TiO_2$/NiFe shows only weak absorption in the visible region. In contrast, $TiO_2$/Au displays a pronounced broad absorption band centered at 580 nm, characteristic of the LSPR of Au nanoparticles. The $TiO_2$/Au/NiFe electrode retains this plasmonic absorption feature, redshifted to 590 nm. This red-shift is consistent with an increased effective refractive index around the Au nanoparticles. The enhanced visible-light absorption of $TiO_2$/Au/NiFe supports its use as a photoanode platform in which Au-generated hot holes can transfer to the NiFe-based OER catalysts while hot electrons are separated at the Au/$TiO_2$ Schottky barrier.[27]

The photoelectrochemical response of the different electrode architectures was evaluated under simulated AM1.5G illumination by chronoamperometry at 1.23 V vs. RHE (**Figure 1d**) and linear sweep voltammetry (LSV, **Figure 1e**). At 1.23 V vs RHE, $TiO_2$/Au/NiFe produced a photocurrent density of 19 $\mu A \cdot cm^{-2}$, compared with 13 $\mu A \cdot cm^{-2}$ for $TiO_2$/NiFe and 10·$\mu A\ cm^{-2}$ for bare $TiO_2$. Although these photocurrent densities are lower than those typically reported for optimized $TiO_2$ photoanodes,[13] the electrodes were intentionally designed for wavelength-resolved photo-SECM measurements rather than maximum performance. In particular, the $TiO_2$ layer was limited to 40 nm to ensure optical transparency for back-side illumination, and the NiFe loading was kept ultrathin to preserve a well-defined interface. The enhanced responses of $TiO_2$/Au and $TiO_2$/NiFe relative to bare $TiO_2$ indicate that both modifications alter the PEC behavior, with Au primarily extending visible-light absorption and NiFe increasing anodic catalytic current at higher potentials. $TiO_2$/NiFe also showed an enhanced dark current at high potentials, consistent with the expected catalytic activity of the NiFe-based species. In addition, $TiO_2$/Au/NiFe showed smaller capacitive effects than $TiO_2$/Au (**Figure 1d**), suggesting improved charge transfer.

The combined $TiO_2$/Au/NiFe architecture exhibited the strongest photocurrent response under AM1.5G illumination and retained a measurable photocurrent under filtered visible-light excitation. In contrast, the photocurrent of $TiO_2$/NiFe became negligible when a 550 nm long-pass filter was used. The persistence of photocurrent under visible-light excitation indicates that Au-mediated plasmonic excitation extends the photoresponse into the visible region, introducing an additional light-harvesting pathway.

The NiFe-containing samples show a more positive photocurrent onset than bare $TiO_2$. This behavior may reflect the potential-dependent formation of oxidized NiFe species, such as Ni(Fe)OOH-like sites, required for efficient OER, but it may also indicate that the NiFe layer modifies interfacial energetics or partially blocks direct charge transfer at low bias.[28] Once sufficiently anodic potentials are reached, the NiFe-containing electrodes exhibit substantially larger photocurrents, consistent with enhanced OER catalysis. Although the apparent onset is shifted positively relative to bare $TiO_2$, it remains below the thermodynamic OER potential of 1.23 V vs RHE under illumination, indicating that the architecture can drive anodic photocurrent at potentials relevant to water oxidation. Notably, as shown by the photo-SECM experiments discussed below, local $O_2$ generation can also be detected under localized illumination while the photoanode is held at open circuit, indicating that this architecture can drive measurable $O_2$ evolution without applying an external bias to the substrate.

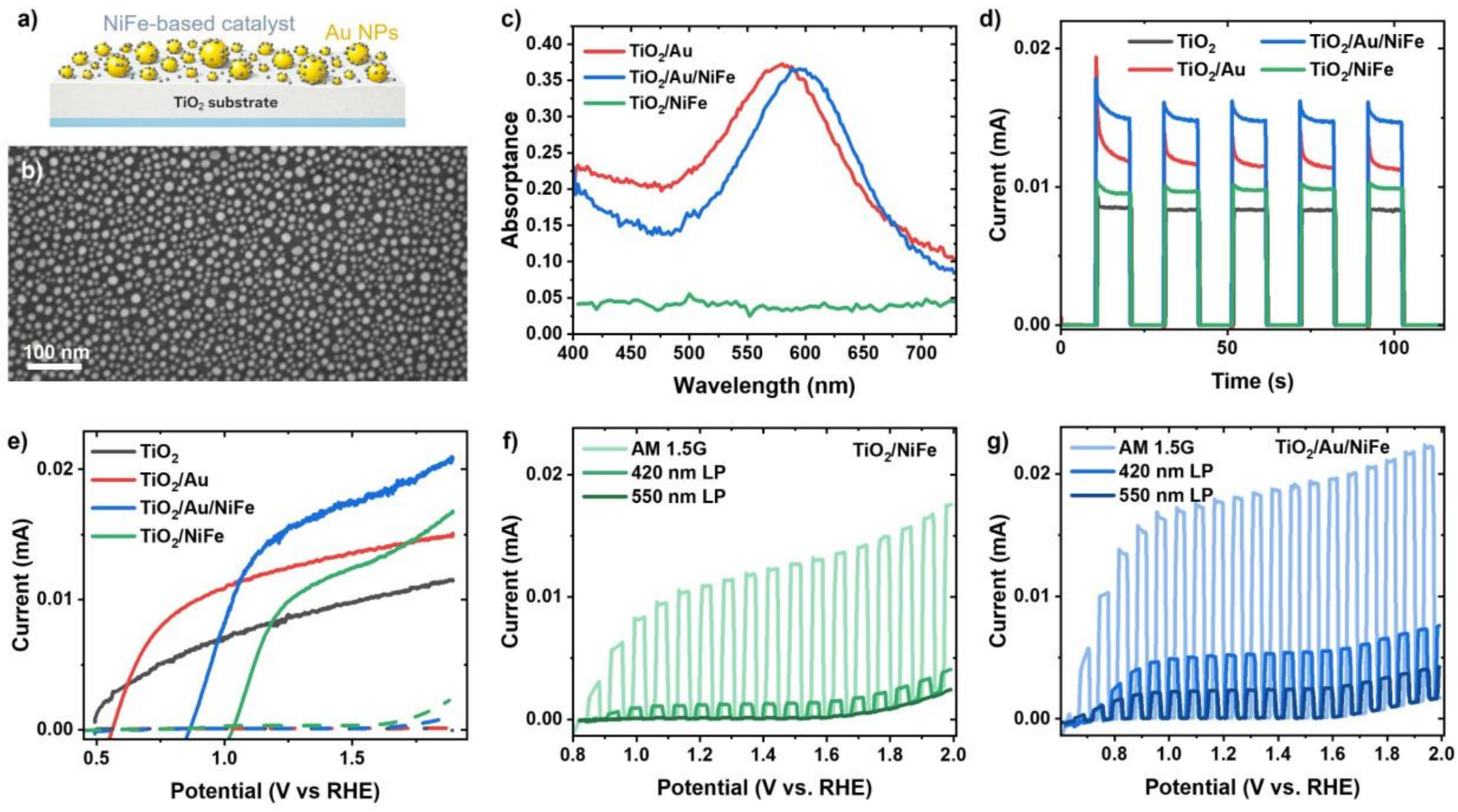

**Figure 1. Photoelectrochemical characterization. a)** Schematic representation of the $TiO_2$/Au/NiFe architecture. **b)** SEM image of the Au nanoparticle–modified $TiO_2$ surface, showing the formation of dispersed Au nanoparticles. **c)** UV–vis absorptance spectra of $TiO_2$/Au, $TiO_2$/NiFe, and $TiO_2$/Au/NiFe electrodes in water. Au–containing samples show a broad visible absorption band assigned to the LSPR of Au nanoparticles, which is retained after NiFe deposition. **d)** Chopped-light chronoamperometry at 1.23 V comparing the PEC response of the different electrode architectures. **e)** Photoelectrochemical response of the electrodes during LSV under AM 1.5G illumination (solid lines) and in the dark (dashed lines). **f,g)** Chopped-light LSV of $TiO_2$/NiFe (f) and $TiO_2$/Au/NiFe (g) under AM1.5G illumination and using different long-pass filters (420 and 550 nm). Scan rate 50 mV·s$^{-1}$, light chopping at 2 Hz.

## 2.2. Photo-SECM detection of local oxygen evolution

Photocurrent measurements provide an initial indication of photoelectrochemical activity, but they do not by themselves confirm that the anodic current originates from water oxidation. In particular, they can include contributions from interfacial charging, parasitic oxidation reactions, or photoelectrode degradation, making direct product detection essential for correlating charge transfer with oxygen evolution.[29] Conventional $O_2$ quantification methods, such as gas chromatography and optical $O_2$ sensors, are widely used for ensemble measurements; however, for wavelength-dependent photoelectrochemical studies, they typically require long times and sufficient product accumulation at each excitation wavelength and can be influenced by $O_2$ diffusion, phase partitioning between electrolyte and headspace, and leakage.[30] Instead, photo-SECM provides an alternative approach by electrochemically detecting $O_2$ as it is generated. This enables operando, localized detection of $O_2$ evolution through picoampere-level oxygen-reduction currents, without requiring macroscopic product accumulation, therefore allowing direct correlations between excitation wavelength, photocurrent response, and $O_2$ generation.[27,31–37]

Photo-SECM was used in the substrate generation/tip collection (SG/TC) mode to directly probe local $O_2$ evolution from the $TiO_2$-based photoanodes under collimated back-side illumination. In this configuration, the photoanode was illuminated with a collimated laser beam (~30 μm diameter) through the transparent substrate, while an ultramicroelectrode (UME) tip positioned above the surface monitored the oxygen generated at the illuminated region (**Figure 2a**). A gold UME with a diameter of approximately 4 μm was positioned 6 μm above the substrate and cathodically biased to reduce locally generated $O_2$. Au UMEs were selected because they exhibited improved stability compared with Pt UMEs under the studied conditions.[38] A detailed comparison between Au and Pt UMEs is provided in **Supplementary Note 1**.

To verify the suitability of the Au UME for oxygen detection, the potential window in which the tip was sensitive to dissolved $O_2$ was first evaluated. Cyclic voltammetries (CVs) were recorded in air-, $N_2$-, and $O_2$-saturated electrolytes, and the cathodic branches are shown in **Figure S5**. In $N_2$-saturated electrolyte, only a small background current was observed over the investigated potential range. In contrast, both air- and $O_2$-saturated electrolytes showed a pronounced increase in cathodic current at negative potentials, consistent with oxygen reduction at the Au UME. The larger current measured in $O_2$- saturated electrolyte compared with air-saturated electrolyte reflects the higher concentration of dissolved $O_2$. Based on these measurements, −1.0 V vs Ag/AgCl was selected as the tip potential for photo-SECM experiments in 0.1 M KOH, providing sufficient sensitivity toward $O_2$ reduction while avoiding significant $H_2$ evolution at more negative potentials.

When the Au UME was positioned close to a $TiO_2$/Au photoanode, the tip current followed the periodic illumination sequence at each excitation wavelength (**Figure 2b**), confirming that the detected signal originated from light-driven $O_2$ formation at the photoanode surface. This effect was not observed when the tip was far from the surface (**Figure S6**). During these measurements, the photoanode was held at 0 V versus its open-circuit potential, and the substrate photocurrent was recorded simultaneously with the UME tip current. This configuration allowed the local $O_2$-detection signal to be directly correlated with the photocurrent response under identical illumination conditions. Analogous tip-current responses were also observed in control measurements where the photoanode was electrically connected as a second working electrode but no potential-control technique was applied to the substrate. This confirms that the detected $O_2$ signal does not require an externally imposed substrate bias under the present localized illumination conditions.

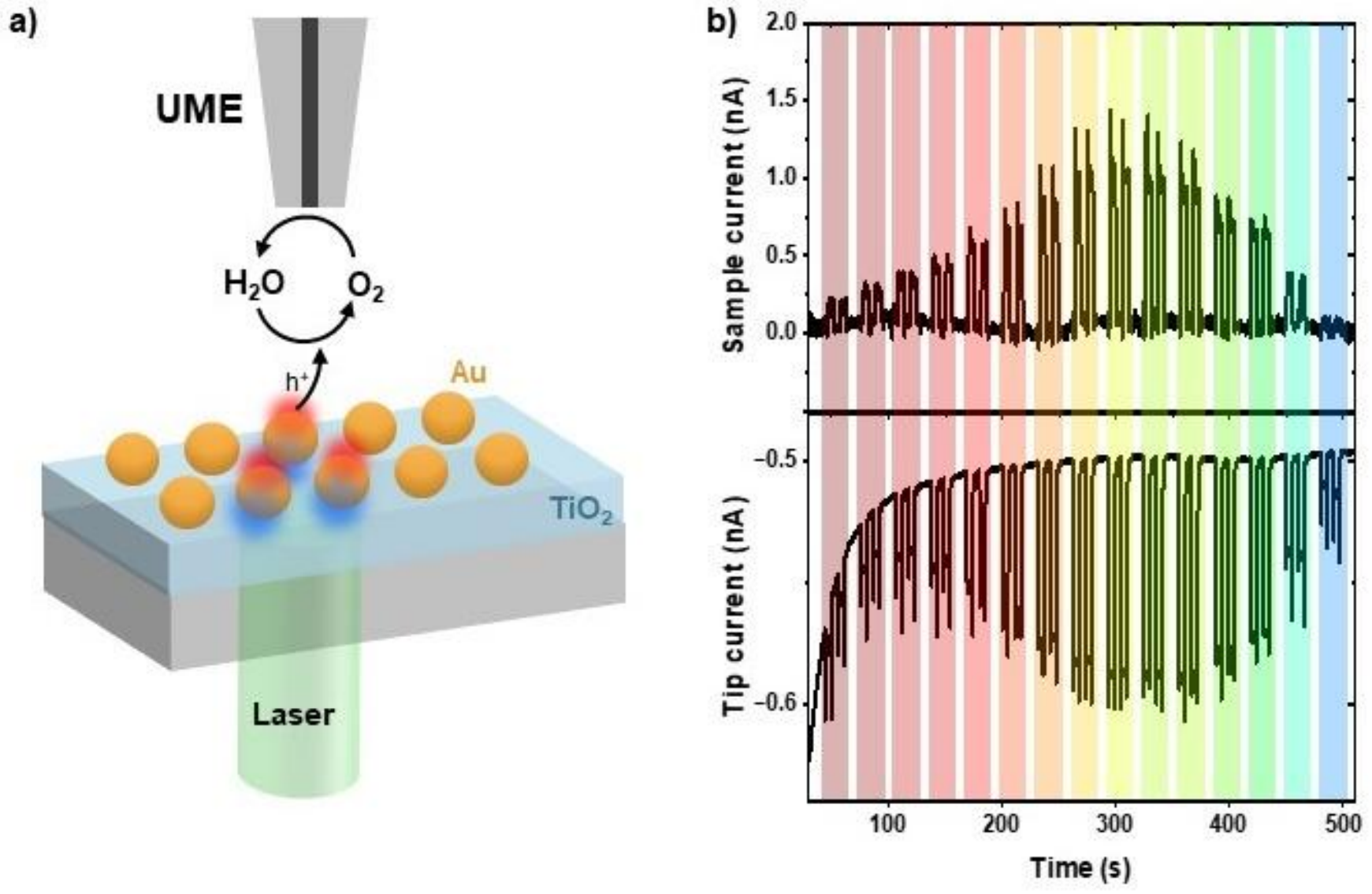

**Figure 2. Local $O_2$ detection via photo-SECM. a)** Schematic representation of the photo-SECM configuration used for local $O_2$ detection under back-side illumination. **b)** Representative sample current and UME tip current recorded simultaneously for a $TiO_2$/Au sample at different excitation wavelengths from 750 to 470 nm, varied in 20 nm

steps. During the measurement, the photoanode was held at its open-circuit potential, while the Au UME was biased at −1.0 V vs Ag/AgCl for $O_2$ reduction.

**Figures 3a,b** show the wavelength-dependent substrate photocurrent and the corresponding tip response, averaged from three on/off cycles (exemplified in **Figure S7**). The substrate photocurrent was calculated as $\Delta i_{sub} = i_{sub,light} - i_{sub,dark}$. The tip response was calculated as $\Delta i_{tip} = i_{tip,light} - i_{tip,dark}$. Since $O_2$ reduction at the tip produces a cathodic current, plotting $-\Delta i_{tip}$ allows larger $O_2$-detection signals to appear as positive values, facilitating a direct comparison with the photocurrent trends.

For bare $TiO_2$, a negligible photocurrent was observed at an incident laser power of 10 μW (1.4 W·cm$^{-2}$) at the studied wavelengths, and $O_2$ detection was only measurable below 500 nm (**Figure S8**). At 1 μW (0.14 W·cm$^{-2}$), neither photocurrent nor $O_2$ generation was detected for bare $TiO_2$, consistent with its limited visible-light response.

In contrast, $TiO_2$ modified with Au nanoparticles showed higher activity across the investigated wavelength range. On these samples, a laser power of 1 μW enabled a reliable comparison of the $O_2$-detection signal at the tip, as discussed in **Supplementary Note 2**. For $TiO_2$/Au, both the photocurrent and the tip response followed the optical absorption profile of the electrode, confirming that Au NPs contribute to light harvesting and charge carrier generation. However, after photo-SECM measurements, optical microscopy revealed a color change in the laser-illuminated region (**Figure S10**), suggesting local degradation or irreversible modification of the $TiO_2$/Au surface under the measurement conditions.

Upon introduction of the NiFe-based catalyst layer, this laser-induced degradation was no longer observed. The $TiO_2$/Au/NiFe electrode showed a similar wavelength-dependent trend in both photocurrent and $O_2$ detection, but the maximum $O_2$-detection signal was shifted toward longer wavelengths, consistent with the red-shifted optical absorption of the full $TiO_2$/Au/NiFe architecture. These results indicate that the NiFe layer stabilizes the plasmonic photoanode and improves the conversion of Au-mediated photoexcitation into oxygen evolution.

To gain further insights into the carrier generation and utilization process, we calculated the internal quantum efficiency (IQE, see details in **Supplementary Note 3**) for $TiO_2$/Au and $TiO_2$/Au/NiFe photoanodes (**Figure 3c**). The $TiO_2$/Au/NiFe electrode exhibits a substantially higher IQE over the entire visible range, reaching ca. $1.5\times10^{-3}$ at 450 nm, whereas $TiO_2$/Au shows a much weaker response that rapidly decays with increasing wavelength. This increase in IQE demonstrates that the NiFe layer enhances the conversion of absorbed plasmonic photons into anodic charge. We attribute this enhancement to the role of NiFe as a catalytic hot-hole extraction layer: upon Au excitation, energetic holes generated in the Au nanoparticles are efficiently transferred to NiFe active sites, where they can accumulate and be consumed by water-oxidation intermediates.[27,39,40] This creates a sink for oxidative carriers, suppressing recombination at the $TiO_2$/Au/electrolyte interface. The blue-shifted IQE maximum relative to the plasmonic absorptance maximum further indicates that photocurrent generation is governed not only by the number of absorbed photons, but by the energy distribution of photoexcited charge carriers per absorbed photon (or at each photon energy). Higher-energy excitation more efficiently generates transferable d-band hot holes in Au due to the interband excitation[27] capable of driving interfacial oxidation, whereas lower-energy carriers are less effective. Therefore, the IQE spectrum provides strong evidence that NiFe promotes harvesting of hot-holes and their conversion into oxygen evolution.

Taken together, the photo-SECM and IQE measurements show that Au nanoparticles extend the $TiO_2$ photoresponse into the visible region and enable $O_2$ evolution at longer wavelengths. NiFe reduces the local degradation observed for $TiO_2$/Au, enhances the absorbed-photon-to-current conversion efficiency, and promotes the utilization of Au-derived hot-holes for $O_2$ evolution. Thus, the $TiO_2$/Au/NiFe architecture functions not only as a plasmonic light absorber, but as an integrated hot-carrier/OER interface.

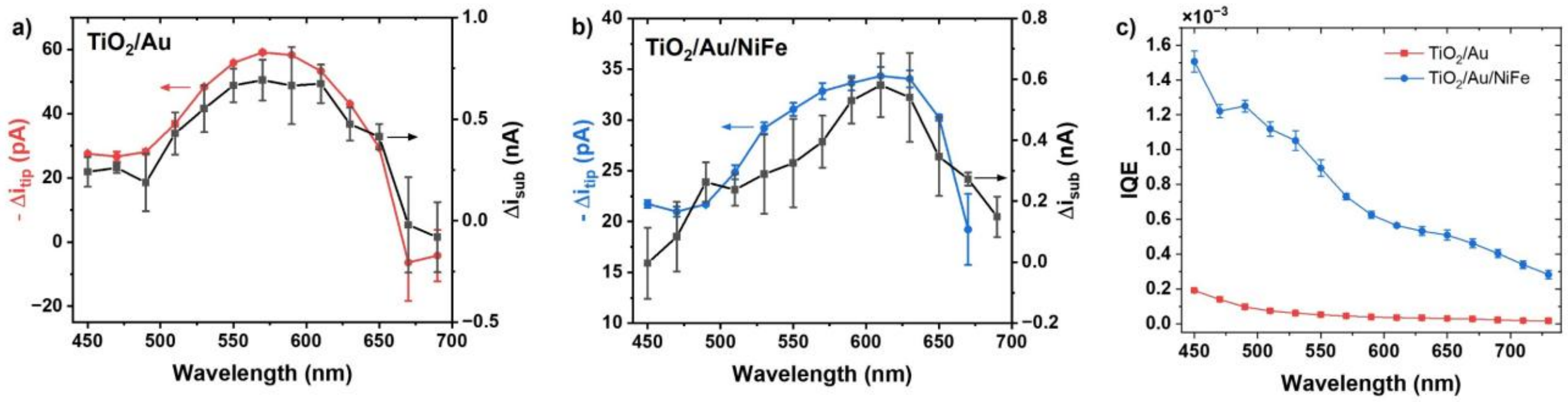

**Figure 3. Wavelength-dependent photoelectrochemical response.** Wavelength-dependent substrate photocurrent ($\Delta i_{sub}$) and corresponding $-\Delta i_{tip}$ values for **a)** $TiO_2$/Au and **b)** $TiO_2$/Au/NiFe. Here, $\Delta i_{tip} = i_{tip,light} - i_{tip,dark}$, and $-\Delta i_{tip}$ is plotted so that larger $O_2$-reduction signals are shown as positive values. **c)** IQE of $TiO_2$/Au and $TiO_2$/Au/NiFe samples.

## 2.3. Chiral molecular functionalization

To introduce spin-polarization at the interface between the plasmonic Au nanoparticles and the NiFe-based catalyst layer via the CISS effect, $TiO_2$/Au electrodes were first functionalized with cysteine, followed by deposition of the NiFe layer. Cysteine was selected because its thiol group can strongly bind to Au, allowing the chiral layer to interact directly with the plasmonic component of the photoanode. L-cysteine (L-cys) was used as a homochiral modifier, while DL-cysteine (DL-cys) acted as a racemic control that provides the same chemical functionality. After functionalization, the samples showed an absorption maximum around 584 nm (**Figure 4a**), slightly blue-shifted relative to the unmodified $TiO_2$/Au/NiFe electrodes, suggesting that cysteine adsorption modifies the local environment or the interaction between Au nanoparticles and the NiFe catalyst layer.

**Figure 4b** shows the photoelectrochemical activity of the cysteine-modified samples. Compared with non-functionalized $TiO_2$/Au/NiFe (shown in **Figure 1**), cysteine-functionalized electrodes presented lower overall photocurrent, suggesting partial passivation of the photoanode interface by the molecular layer. This decrease may arise from reduced electronic coupling between Au and NiFe, modification of the local plasmonic environment, or partial blocking of OER-active NiFe sites.

Despite this passivation, the L-cys-functionalized electrode showed a higher photocurrent than the DL-cys-functionalized control. Under simulated solar illumination, L-cys shifted the apparent photocurrent onset to lower potentials and increased the photocurrent over the investigated potential range. The dark current at high anodic bias was also increased, consistent with previous reports on similar chiral-modified OER systems, where chiral molecular layers strongly influence electrocatalytic water oxidation in the dark.[9] The enhanced photocurrent of the L-cys-modified sample points to more favorable interfacial charge-transfer kinetics and a lower energetic requirement for oxygen evolution, consistent with a contribution from the CISS effect.

The difference became more pronounced under 550 nm long-pass illumination (**Figure 4c**), where direct $TiO_2$ band-gap excitation is suppressed and the response is dominated by Au-mediated excitation. At 1.23 V vs RHE under AM1.5G illumination, $TiO_2$/Au/L-cys/NiFe produced 14.2 $\mu A \cdot cm^{-2}$, compared with 13.1 $\mu A \cdot cm^{-2}$ for $TiO_2$/Au/DL-cys/NiFe, corresponding to an 8% enhancement. Under 550 nm long-pass filtered illumination, the enhancement increased to 130%. This result suggests that the photocurrent generated through LSPR excitation is particularly sensitive to the presence of the chiral molecular layer. Importantly, the DL-cys-modified electrode exhibited slightly higher absorptance than the L-cys-modified electrode in the visible region (**Figure 4a**), excluding stronger light absorption as the origin of the enhanced L-cys photocurrent.

Wavelength-dependent photo-SECM measurements were then used to determine whether the chirality-dependent photocurrent enhancement also corresponded to increased $O_2$ generation (**Figure 4d**). Compared with DL-cys, the L-cys-modified electrode showed enhanced local $O_2$ detection, with the largest difference near the plasmonic absorption region. The increase in tip current upon illumination ($\Delta i_{tip,L\text{-}cys} - \Delta i_{tip,DL\text{-}cys}$) was 0.5 pA at 450 nm, 1.3 pA at 590 nm, and 0.05 pA at 690 nm. The maximum enhancement near the Au plasmon band indicates that homochiral functionalization improves the conversion of plasmon-assisted excitation into $O_2$ formation.

The substrate photocurrent spectra measured at higher incident intensity (**Figure 4e**) show the same trend, with L-cys producing higher photocurrent than DL-cys over the spectral region where the $O_2$-detection enhancement is most pronounced. In **Figure 4f**, L-cys-modified photoanode exhibits higher IQE than the DL-cys control over most of the visible range, indicating that the homochiral interface increases the fraction of absorbed photons converted into anodic charge. Because IQE is normalized by absorbed photon flux, this enhancement reflects improved carrier utilization rather than differences in light harvesting.

Comparison with the unmodified $TiO_2$/Au/NiFe electrode further clarifies the role of the molecular layer. Direct contact between Au and NiFe gives the highest IQE, consistent with efficient extraction and consumption of Au-derived hot holes at NiFe active sites. Inserting cysteine lowers the IQE, indicating that the molecule partially reduces direct Au–NiFe electronic coupling and introduces an additional transport barrier. However, within this molecularly modified interface, L-cys consistently outperforms DL-cys. This shows that although cysteine partially passivates the interface, homochiral cysteine could be enabling spin-selective hot-hole transfer through the Au–cysteine–NiFe junction.

Overall, the higher photocurrent, local $O_2$ detection, and IQE of $TiO_2$/Au/L-cys/NiFe relative to $TiO_2$/Au/DL-cys/NiFe are consistent with a CISS-related contribution, in which the chiral molecular layer modulates spin-dependent hot-hole transfer at the Au–molecule–NiFe interface.

This interpretation also helps place the present results in the context of recent studies on chiral plasmonic nanoparticles for spin-selective oxygen electrocatalysis,[19,21] where chiral ligands or templates used during synthesis can couple nanoparticle structural chirality with surface-molecular chirality. The present architecture uses structurally achiral Au nanoparticles and introduces chirality only through the cysteine interface, allowing the molecular contribution to plasmon-derived spin-selective charge transfer to be isolated. These results therefore suggest that structural chirality and surface-molecular chirality may serve as complementary knobs for controlling spin-selective plasmon-assisted catalysis.

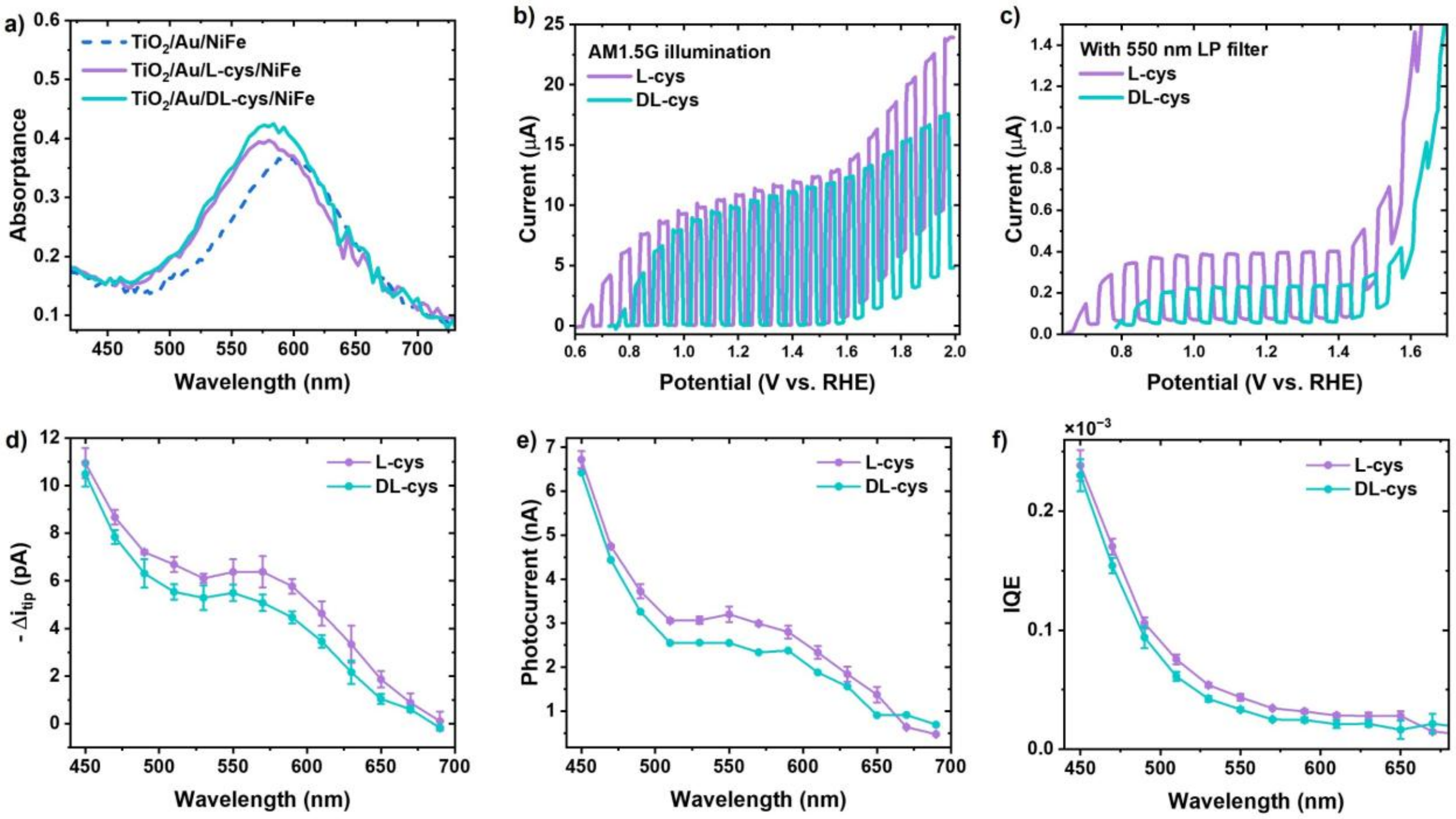

**Figure 4. Chiral molecule modifications. a)** Absorptance spectra of unmodified $TiO_2$/Au/NiFe and cysteine-modified samples in water. **b)** PEC LSV with chopped AM1.5G illumination and **c)** a 550 nm long-pass filter of samples modified with L-cysteine (homochiral) and DL-cysteine (racemic). **d)** Wavelength-dependent $O_2$-detection signal, plotted as $-\Delta i_{tip}$, for cysteine-modified photoanodes at 1.4 W·cm$^{-2}$ incident intensity. **e)** Substrate photocurrent measured at 70 W·cm$^{-2}$ incident intensity. **f)** IQE calculated from the absorptance and photocurrent spectra.

Based on these findings, we propose the mechanism schematized in **Figure 5**. Under visible-light illumination, excitation of the Au LSPR generates energetic hot-carriers at the Au/$TiO_2$ interface. Hot electrons can be injected into the $TiO_2$ conduction band, while the corresponding hot holes are transferred toward the NiFe oxygen-evolution catalyst. When a homochiral cysteine layer is present between the Au nanoparticles and the catalyst, this interfacial hole transfer can become spin-polarized through the CISS effect. This can favor spin-allowed pathways for the formation of triplet $O_2$.

Further optimization of molecular coverage, anchoring geometry, and catalyst accessibility will be required to maximize the spin-selective enhancement while minimizing interfacial passivation.

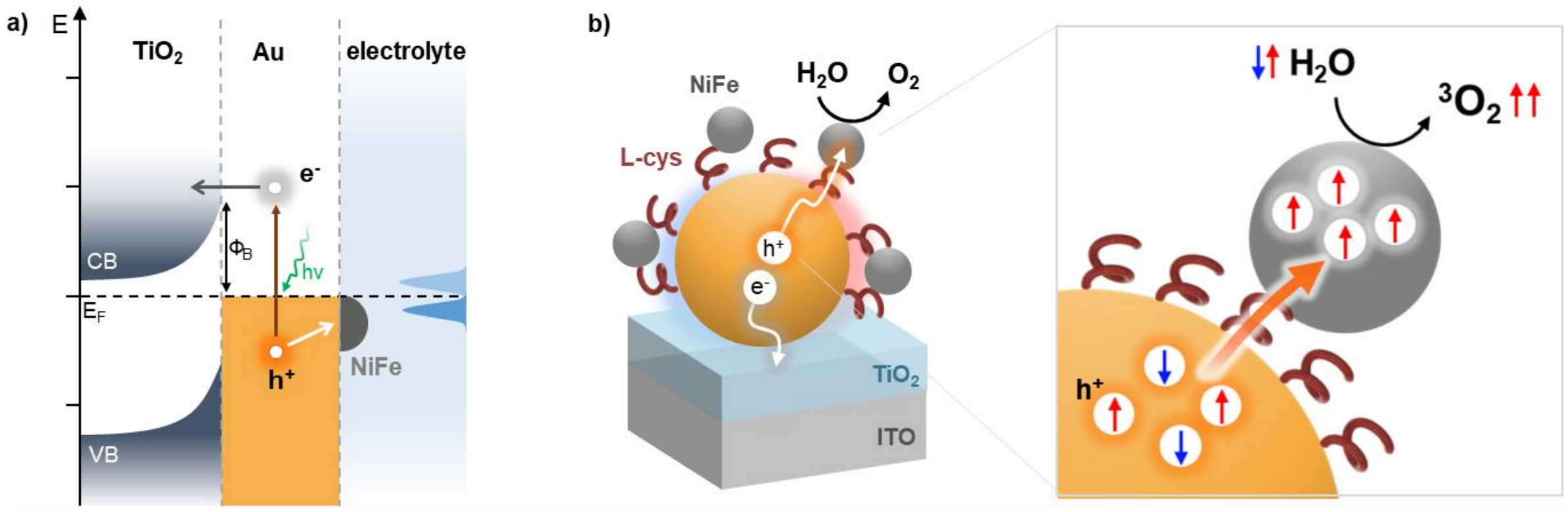

**Figure 5. Proposed mechanism for spin-selective plasmon-assisted oxygen evolution at $TiO_2$/Au/L-cys/NiFe photoanodes. a)** Schematic band-alignment diagram of the $TiO_2$/Au interface, showing the $TiO_2$ valence and conduction bands (VB and CB, respectively) and the Schottky barrier formed at the Au/$TiO_2$ junction. Visible-light excitation of the Au LSPR generates hot-carriers: hot electrons are injected into the $TiO_2$ conduction band, while hot holes are transferred toward NiFe OER sites, where they participate in water oxidation. **b)** Proposed CISS-mediated hot hole transfer through the homochiral L-cys interface in an unbiased condition. The cysteine layer can spin-polarize the transferred hot holes, increasing the population of spin-polarized oxidative equivalents at NiFe sites and facilitating spin-allowed triplet oxygen formation.

## Conclusions

In this work, we developed a hybrid plasmonic photoanode based on $TiO_2$ modified with Au nanoparticles and a NiFe-based oxygen evolution catalyst. The incorporation of Au nanoparticles introduced a visible-light absorption pathway associated with localized surface plasmon resonance, while the NiFe layer improved the catalytic activity of the plasmonic interface under photoelectrochemical operation.

Photo-SECM measurements enabled direct, local detection of oxygen evolution from the illuminated photoanode surface. The wavelength-dependent $O_2$-detection signal correlated with the optical absorption and photocurrent response of the plasmonic electrodes, confirming that Au nanoparticles enhance light-driven oxygen evolution beyond the intrinsic UV response of $TiO_2$. Importantly, the NiFe-based catalyst layer reduced the laser-induced degradation observed for $TiO_2$/Au electrodes, indicating that the catalyst layer not only facilitates OER but also stabilizes the plasmonic photoanode during illumination.

Chiral molecular functionalization with L-cysteine and racemic DL-cysteine provided a platform to examine the CISS effect at the plasmonic nanoparticle–molecule–catalyst interface. Although cysteine adsorption partially passivated the electrode compared to the unmodified $TiO_2$/Au/NiFe system, the homochiral L-cysteine layer produced higher photocurrent and local $O_2$ evolution than the racemic control, particularly under visible-light excitation overlapping with the Au plasmonic absorption. This chirality-dependent response suggests that the molecular layer influences hot-carrier transfer, consistent with spin-polarized charge transfer at the interface. More broadly, these results complement previous studies on chiral plasmonic nanoparticles, which often employ chiral ligands during synthesis or stabilization, and suggest that both structural and surface-molecular chirality may act together in spin-selective plasmon-assisted catalysis.

Overall, we present a chiral plasmonic photoelectrochemical platform in which achiral Au nanoparticles, chiral molecular layers, and NiFe OER catalysis are coupled within a single architecture. The observed chirality-dependent oxygen evolution represents the first demonstration of spin-polarization of hot-carriers in a hybrid achiral plasmonic NP-chiral molecule-catalyst system. While further optimization is needed to minimize molecular passivation and improve catalytic accessibility, this work establishes a foundation for using chiral molecular interfaces to control plasmon-assisted water oxidation and spin-dependent photoelectrochemical processes.

## Methods

*Electrode preparation*

Transparent conductive substrates were prepared by depositing a 50 nm ITO layer by RF sputtering using a Pfeiffer SPIDER 600 system onto 520 µm-thick fused silica substrates. Subsequently, a 40 nm $TiO_2$ layer was deposited by sputtering using an Alliance-Concept DP 650 system. The resulting $TiO_2$/ITO films were annealed in air at 450 °C for 2 h and then diced into 1.7 × 1.7 $cm^2$ pieces.

Au nanoparticles were formed by sputtering a nominal 2 nm Au layer onto the $TiO_2$/ITO substrates using an Alliance-Concept DP 650 system, followed by annealing in air at 300 °C for 2.5 h to induce dewetting of the thin Au film and formation of Au nanoparticles. Finally, a nominal 0.5 nm NiFe (Ni/Fe 81/19 wt. %) layer was deposited by e-beam evaporation using a Leybold Optics LAB 600H system to introduce a thin OER catalyst layer. The resulting samples are denoted $TiO_2$/Au/NiFe.

For chiral molecular functionalization, a 10 $mg \cdot ml^{-1}$ aqueous solution of the corresponding cysteine was drop-casted onto the $TiO_2$/Au surface and allowed to incubate for 30 min. The samples were then thoroughly rinsed with water to remove weakly adsorbed molecules and dried under $N_2$ flow. The NiFe catalyst layer was then deposited by evaporation as described above. For photo-SECM measurements, a single $TiO_2$/Au substrate was spatially divided into two regions with Kapton tape. One half of the substrate was functionalized with L-cysteine, while the other half was functionalized with racemic DL-cysteine. After both regions had been functionalized, the substrate was thoroughly rinsed with water, dried under $N_2$ flow, and coated uniformly with the NiFe catalyst layer.

*Structural and optical characterization*

Absorptance spectra were acquired in a liquid compression cell filled with water, with light incident through the fused silica/ITO substrate. Reflectance (R) and transmittance (T) spectra were measured using an integrating sphere (Newport, 819D-IS-5.3) coupled to a monochromator (Newport, CS260B-2-MC-A), with a solar simulator (Newport 66984-300XF-R1 Xe lamp) equipped with an AM 1.5G filter as the light source. The absorptance was calculated as: $A = 1 - R - T$. [41]

FE-SEM images were obtained using a Zeiss Merlin microscope at the Interdisciplinary Centre for Electron Microscopy at EPFL (CIME, EPFL). Image analysis was performed using Fiji/ImageJ to determine the Au nanoparticle size distribution.

*Photoelectrochemical measurements*

Sample-scale photoelectrochemical measurements were performed in a three-electrode configuration using the modified $TiO_2$ photoanode as the working electrode, Au foil as the counter electrode, and a Reference Hydrogen Electrode (RHE, Gaskatel). Measurements were recorded using a BioLogic SP-300 potentiostat. The electrolyte was 0.1 M KOH.

Illumination was provided by a solar simulator equipped with a Newport 66984-300XF-R1 Xe lamp and an AM 1.5G filter under 1 sun illumination calibrated to 100 mW $cm^{-2}$ at the sample position. Additional wavelength-selective measurements were performed using long-pass filters at 420 nm and 550 nm (Thorlabs FGL420S and FGL550S). All samples were back-illuminated through the ITO/fused silica substrate, with an illuminated area of 0.785 $cm^2$.

*Scanning photo-electrochemical microscopy*

Photo-SECM was used to locally detect oxygen formation using a previously described home-built instrument.[27] Briefly, a custom-built SECM setup was integrated with an inverted optical microscope (Nikon Eclipse Ti2) and a high-power supercontinuum white-light laser (NKT Photonics), which provided bottom illumination through the transparent substrate. The excitation beam was collimated to a diameter of approximately 30 μm at the sample plane. A tunable wavelength filter (SuperK VARIA,

NKT Photonics) was used to select the excitation wavelength and power over the 450–750 nm spectral range with a bandwidth of 10 nm.

Au UMEs of ~4 μm diameter were used as SECM tips for oxygen-detection measurements because they provided a more stable oxygen-reduction current than Pt UMEs under the present photo-SECM conditions, as detailed in **Supplementary Note 1**. The tip was held at -1 V vs Ag/AgCl during oxygen-detection experiments. The photoanode substrate was kept at 0 V vs OCP, and the tip-substrate distance was set to 6 μm. A Pt wire and a leak-free Ag/AgCl electrode (LF-1, 1 mm OD, Innovation Instrument) were used as counter and reference electrodes, respectively.

For wavelength-dependent photo-SECM measurements, the substrate was illuminated sequentially at selected wavelengths between 450 and 710 nm in 20 nm intervals. The optical power was adjusted to the target value at each wavelength, e.g. 1 or 10 μW, and calibrated using two power meters, one on the sample plane (Thorlabs S170C) and one in the beam path (Thorlabs S130C) to control the power during photo-SECM experiments. At each wavelength, three consecutive illumination cycles were performed using a mechanical shutter (Thorlabs SC10) while monitoring the SECM tip current for $O_2$ detection in SG/TC mode and the substrate current simultaneously.

For each illumination cycle, the light-induced change in tip current was calculated as $\Delta i_{tip} = i_{tip,light} - i_{tip,dark}$, where $i_{tip,dark}$ corresponds to the current measured during the light-off period immediately preceding illumination. Specifically, the last N data points of each light-off and light-on period were averaged to obtain $i_{tip,dark}$ and $i_{tip,light}$, respectively, with N = 10 or 15 depending on the experiment. The same time windows were used to calculate the corresponding substrate-current response as $\Delta i_{sub} = i_{sub,light} - i_{sub,dark}$. The reported $\Delta i_{tip}$ and $\Delta i_{sub}$ values were obtained by averaging the three on–off differences at each wavelength, and the error bars represent the corresponding standard deviations. A representative example of this analysis procedure is shown in **Figure S7**, which was performed with a Python code. Because only the localized laser spot with a diameter of approximately 30 μm was illuminated, the absolute substrate photocurrents measured in these photo-SECM experiments were small compared with those obtained under full-area illumination.

Slight variations in the absolute UME response were observed between different tips and on different measurement days. Therefore, wavelength-dependent photo-SECM data were analyzed using relative changes in the detected tip current measured under internally consistent conditions for each sample. Direct comparison of absolute tip-current values between different samples was thus avoided.

For the cysteine-modified samples, the L-cys- and DL-cys-functionalized regions of the same split substrate were measured using the same Au UME within the same experimental session and under otherwise identical conditions. To minimize potential artifacts associated with measurement order or temporal drift, measurements were performed in both sequences: L-cys followed by DL-cys, and DL-cys followed by L-cys. Fresh, previously unilluminated locations were used for each measurement.

Wavelength-dependent internal quantum efficiency (IQE) values were calculated from the photocurrent response and the optical absorptance of the photoanodes. Further details of the calculations are given in **Supplementary Note 3**.

## Acknowledgements

P.V., M.S., F.K., G.G.M., and G.T. acknowledge financial support from SNSF Eccellenza Grant PCEGP2-194181. P.V. gratefully acknowledges Elif Nur Dayi for assistance with absorption spectroscopy data analysis, Prof. Haussener for access to the spectrophotometer, and F. A. Rovner for fabricating the automated shutter used in chopped-illumination photoelectrochemical measurements. The authors also acknowledge the support of the following experimental facilities at EPFL: Centre for MicroNanofabrication (CMi) and Interdisciplinary Centre for Electron Microscopy (CIME).

## Author contributions

P.V. and G.T. conceived the project. P.V. led the investigation, including fabrication, optical characterization, SEM imaging, photoelectrochemical measurements, photo-SECM experiments, data analysis, and figure preparation. P.V. and M.S. fabricated the UME tips. F.K. and M.S. developed the photo-SECM setup. G.G.M. developed the software for wavelength and power control. P.V. wrote the manuscript with input from all authors. G.T. supervised the project and acquired funding.

## References

1. Tagliabue, G., Atwater, H. A., Polman, A. & Cortés, E. Photonic solutions help fight climate crisis. *Nat. Photon.* **18**, 879–882 (2024).
2. *Photoelectrochemical Solar Fuel Production*. (Springer International Publishing, Cham, 2016). doi:10.1007/978-3-319-29641-8.

3. Sivula, K. & van de Krol, R. Semiconducting materials for photoelectrochemical energy conversion. *Nat Rev Mater* **1**, 1–16 (2016).
4. van der Minne, E. *et al.* Spin Matters: A Multidisciplinary Roadmap to Understanding Spin Effects in Oxygen Evolution Reaction During Water Electrolysis. *Advanced Energy Materials* e03556 (2025) doi:10.1002/aenm.202503556.
5. Yu, A., Zhang, Y., Zhu, S., Wu, T. & Xu, Z. J. Spin-related and non-spin-related effects of magnetic fields on water oxidation. *Nat Energy* **10**, 435–447 (2025).
6. Naaman, R., Paltiel, Y. & Waldeck, D. H. Chiral Induced Spin Selectivity Gives a New Twist on Spin-Control in Chemistry. *Acc. Chem. Res.* **53**, 2659–2667 (2020).
7. Bloom, B. P., Paltiel, Y., Naaman, R. & Waldeck, D. H. Chiral Induced Spin Selectivity. *Chem. Rev.* **124**, 1950–1991 (2024).
8. Vadakkayil, A. *et al.* Chiral electrocatalysts eclipse water splitting metrics through spin control. *Nat Commun* **14**, 1067 (2023).
9. Liang, Y. *et al.* Enhancement of electrocatalytic oxygen evolution by chiral molecular functionalization of hybrid 2D electrodes. *Nat Commun* **13**, 3356 (2022).
10. Vensaus, P. *et al.* Hybrid mesoporous electrodes evidence CISS effect on water oxidation. *The Journal of Chemical Physics* **160**, 111103 (2024).
11. Zhang, W., Banerjee-Ghosh, K., Tassinari, F. & Naaman, R. Enhanced Electrochemical Water Splitting with Chiral Molecule-Coated Fe3O4 Nanoparticles. *ACS Energy Lett.* **3**, 2308–2313 (2018).
12. Mtangi, W., Kiran, V., Fontanesi, C. & Naaman, R. Role of the Electron Spin Polarization in Water Splitting. *J. Phys. Chem. Lett.* **6**, 4916–4922 (2015).
13. Zhang, X. *et al.* Recent Advances in TiO2-based Photoanodes for Photoelectrochemical Water Splitting. *Chemistry – An Asian Journal* **17**, e202200668 (2022).
14. Tian, Y. & Tatsuma, T. Mechanisms and Applications of Plasmon-Induced Charge Separation at TiO2 Films Loaded with Gold Nanoparticles. *J. Am. Chem. Soc.* **127**, 7632–7637 (2005).
15. Tian, Y. & Tatsuma, T. Plasmon-induced photoelectrochemistry at metal nanoparticles supported on nanoporous TiO2. *Chem. Commun.* 1810–1811 (2004) doi:10.1039/B405061D.
16. Woo Moon, C., Choi, M.-J., Kartham Hyun, J. & Won Jang, H. Enhancing photoelectrochemical water splitting with plasmonic Au nanoparticles. *Nanoscale Advances* **3**, 5981–6006 (2021).
17. Ezendam, S. *et al.* Hybrid Plasmonic Nanomaterials for Hydrogen Generation and Carbon Dioxide Reduction. *ACS Energy Lett.* **7**, 778–815 (2022).
18. Nishijima, Y. *et al.* Near-Infrared Plasmon-Assisted Water Oxidation. *J. Phys. Chem. Lett.* **3**, 1248–1252 (2012).

19. Wang, Z. *et al.* Boosting the Selectivity in Oxygen Electrocatalysis Using Chiral Nanoparticles as Electron-Spin Filters. *J. Am. Chem. Soc.* https://doi.org/10.1021/jacs.5c03394 (2025) doi:10.1021/jacs.5c03394.
20. Chae, K. *et al.* Spin-polarized Acidic Water Electrolysis with Antenna-Reactor Plasmonic Electrocatalysts. *Advanced Materials* **37**, 2507658 (2025).
21. Zhou, H. *et al.* Spin-Polarization Tailored Oxygen Evolution of Chiral Gold Nanocrystals Probed by Single-Particle Electrochemistry. *ACS Nano* **20**, 15697–15705 (2026).
22. Dionigi, F. & Strasser, P. NiFe-Based (Oxy)hydroxide Catalysts for Oxygen Evolution Reaction in Non-Acidic Electrolytes. *Advanced Energy Materials* **6**, 1600621 (2016).
23. *Scanning Electrochemical Microscopy*. (CRC Press, Boca Raton, 2012). doi:10.1201/b11850.
24. Vensaus, P. *et al.* Ni nanoclusters as oxygen evolution catalysts on porous supports for electro- and photocatalysis. *Nanotechnology* **37**, 045401 (2026).
25. Trotochaud, L., Young, S. L., Ranney, J. K. & Boettcher, S. W. Nickel–Iron Oxyhydroxide Oxygen-Evolution Electrocatalysts: The Role of Intentional and Incidental Iron Incorporation. *J. Am. Chem. Soc.* **136**, 6744–6753 (2014).
26. Friebel, D. *et al.* Identification of Highly Active Fe Sites in (Ni,Fe)OOH for Electrocatalytic Water Splitting. *J. Am. Chem. Soc.* **137**, 1305–1313 (2015).
27. Kiani, F. *et al.* Transport and Interfacial Injection of d-Band Hot Holes Control Plasmonic Chemistry. *ACS Energy Lett.* **8**, 4242–4250 (2023).
28. Vensaus, P. *et al.* CoPi-Modified Mesoporous Titania Photoelectrodes for Water Splitting: Why Less Is More. *ACS Appl. Eng. Mater.* **2**, 224–235 (2024).
29. Khan, M. A. *et al.* Importance of Oxygen Measurements during Photoelectrochemical Water-Splitting Reactions. *ACS Energy Lett.* **4**, 2712–2718 (2019).
30. Tiwari, C. K. & Geletii, Y. V. Measurements of Dioxygen Formation in Catalytic Electrochemical Water Splitting. *Catalysts* **14**, 13 (2024).
31. Ye, H., Park, H. S. & Bard, A. J. Screening of Electrocatalysts for Photoelectrochemical Water Oxidation on W-Doped $BiVO_4$ Photocatalysts by Scanning Electrochemical Microscopy. *J. Phys. Chem. C* **115**, 12464–12470 (2011).
32. Chen, X., Botz, A. J. R., Masa, J. & Schuhmann, W. Characterisation of bifunctional electrocatalysts for oxygen reduction and evolution by means of SECM. *J Solid State Electrochem* **20**, 1019–1027 (2016).
33. Minguzzi, A., Alpuche-Aviles, M. A., López, J. R., Rondinini, S. & Bard, A. J. Screening of Oxygen Evolution Electrocatalysts by Scanning Electrochemical Microscopy Using a Shielded Tip Approach. *Anal. Chem.* **80**, 4055–4064 (2008).

34. Bae, J. H., Nepomnyashchii, A. B., Wang, X., Potapenko, D. V. & Mirkin, M. V. Photo-Scanning Electrochemical Microscopy on the Nanoscale with Through-Tip Illumination. *Anal. Chem.* **91**, 12601–12605 (2019).
35. Kiani, F., Bowman, A. R., Sabzehparvar, M., Sundararaman, R. & Tagliabue, G. Distinguishing Inner and Outer-Sphere Hot Electron Transfer in Au/p-GaN Photocathodes. *Nano Lett.* **24**, 16008–16014 (2024).
36. Kiani, F. *et al.* Revealing Wavelength- and Size-Dependent CO2 Reduction Selectivity via Operando Scanning Photo-Electrochemical Microscopy. Preprint at https://doi.org/10.48550/arXiv.2606.09612 (2026).
37. Sabzehparvar, M. *et al.* Thin-Wall Single-Crystal Gold Nanoelectrodes toward Advanced Chemical Probing and Imaging. *Small* **22**, e14938 (2026).
38. Limani, N. *et al.* Low loading ORR selectivity evaluation of Pt-free catalysts with scanning electrochemical microscopy. *Electrochemistry Communications* **153**, 107538 (2023).
39. Laskowski, F. A. L., Nellist, M. R., Qiu, J. & Boettcher, S. W. Metal Oxide/(oxy)hydroxide Overlayers as Hole Collectors and Oxygen-Evolution Catalysts on Water-Splitting Photoanodes. *J. Am. Chem. Soc.* **141**, 1394–1405 (2019).
40. Yu, Y., Wijesekara, K. D., Xi, X. & Willets, K. A. Quantifying Wavelength-Dependent Plasmonic Hot Carrier Energy Distributions at Metal/Semiconductor Interfaces. *ACS Nano* **13**, 3629–3637 (2019).
41. Bowman, A. R., Ma, J., Kiani, F., García Martínez, G. & Tagliabue, G. Best practices in measuring absorption at the macro- and microscale. *APL Photonics* **9**, 061101 (2024).

# Supplementary Material

# Spin-Polarized Oxygen Evolution in Chiral-Molecule-Modified Plasmonic Photoanodes

Priscila Vensaus[1], Milad Sabzehparvar[1], Fatemeh Kiani[1], Germán García Martínez[1], Giulia Tagliabue[1*]

1) Laboratory of Nanoscience for Energy Technologies (LNET), STI, École Polytechnique Fédérale de Lausanne, 1015 Lausanne, Switzerland

*E-mail: giulia.tagliabue@epfl.ch

## Contents

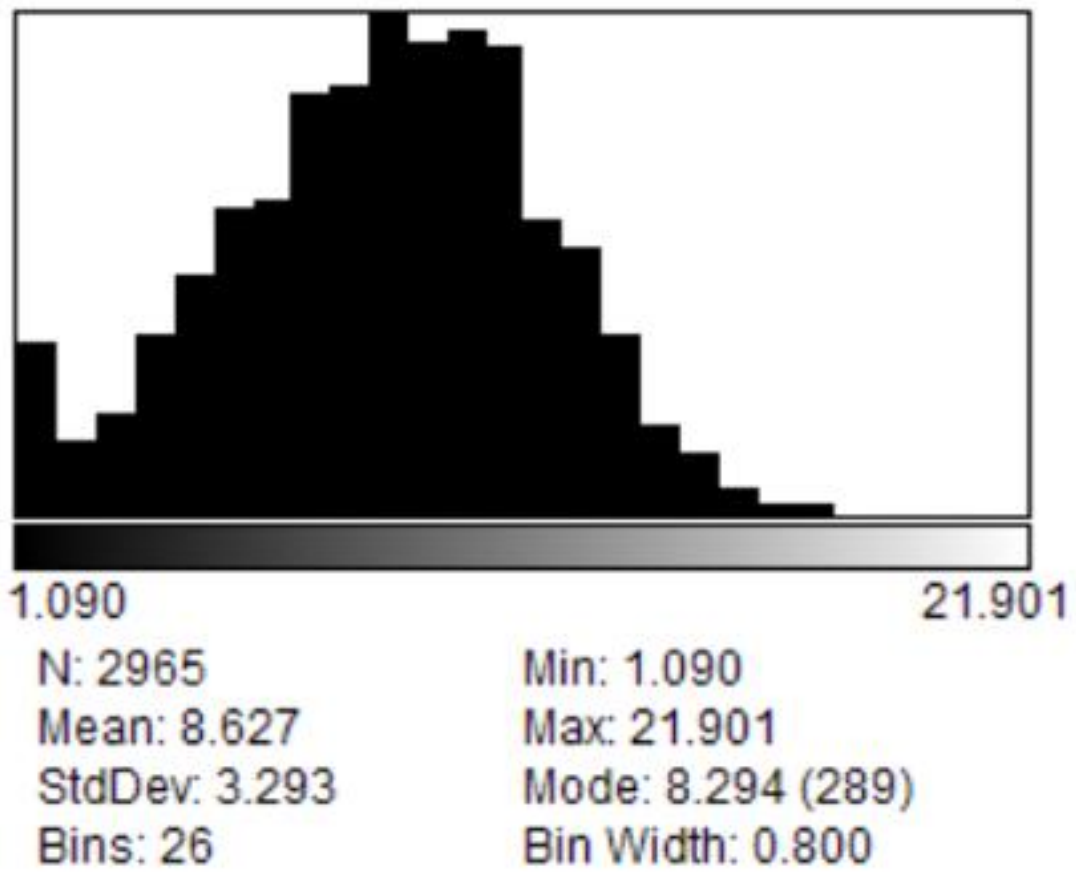

**Figure S1.** Au nanoparticle size distribution, obtained by image analysis with Fiji/ImageJ.

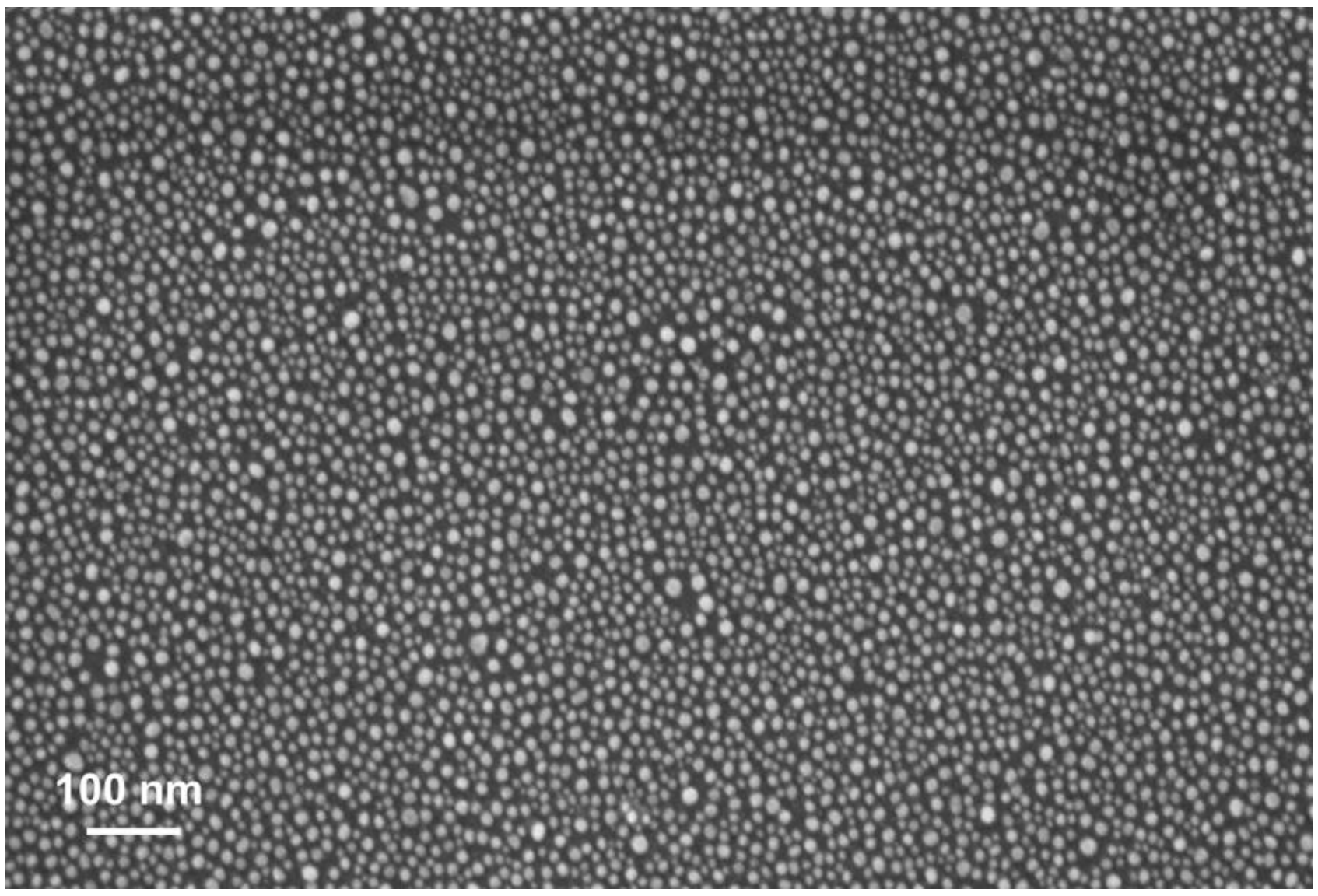

**Figure S2.** SEM image of NiFe-modified $TiO_2$/Au. NiFe deposition does not cause obvious morphology changes of the Au NPs.

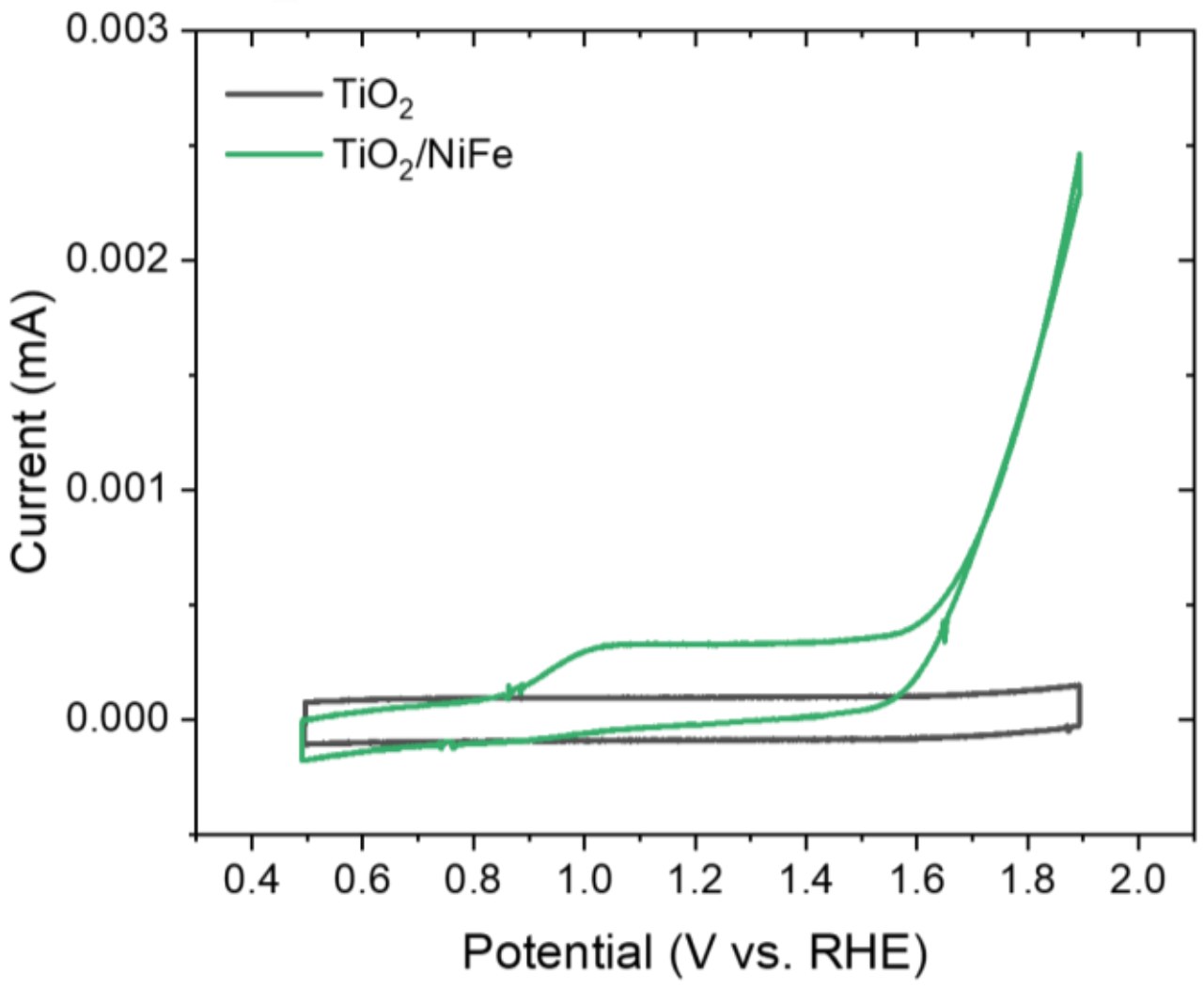

**Supplementary Figure S3.** Dark electrochemical response of $TiO_2$/NiFe, showing the $Ni^{2+}/Ni^{3+}$ redox feature at 1.05 V vs. RHE.

## Supplementary Note 1. Comparison of Au and Pt UMEs for stable oxygen detection in photo-SECM

To identify a suitable SECM tip material for monitoring generated oxygen, Pt and Au UMEs were tested for photo-SECM conditions. Both materials are commonly used for electrochemical oxygen reduction; however, stable tip response is essential for wavelength-dependent measurements, where multiple illumination conditions are compared sequentially. Any drift in the oxygen-reduction current can introduce apparent changes in local $O_2$ flux that are unrelated to the photoanode activity.

### *Preparation of Pt and Au ultramicroelectrode tips*

Pt and Au ultramicroelectrodes were prepared using the same general procedure. Briefly, metal microwires of 25 μm (Goodfellow) were sealed in borosilicate glass capillaries (1 mm OD, 0.5 mm ID, Sutter Instruments), heat-shrunk and pulled using a laser puller (P-2000, Sutter Instruments). Physical contacting of a Cu wire to the Pt/Au wire was done using a silver epoxy. The sealed electrodes were polished using an ultrafine polishing plate (BV-10, Sutter Instruments) to expose a clean disk-shaped metal surface. After polishing, the tips were rinsed thoroughly with ultrapure water, and dried under $N_2$ flow. Figure S4 shows representative microscopy images of an Au UME.

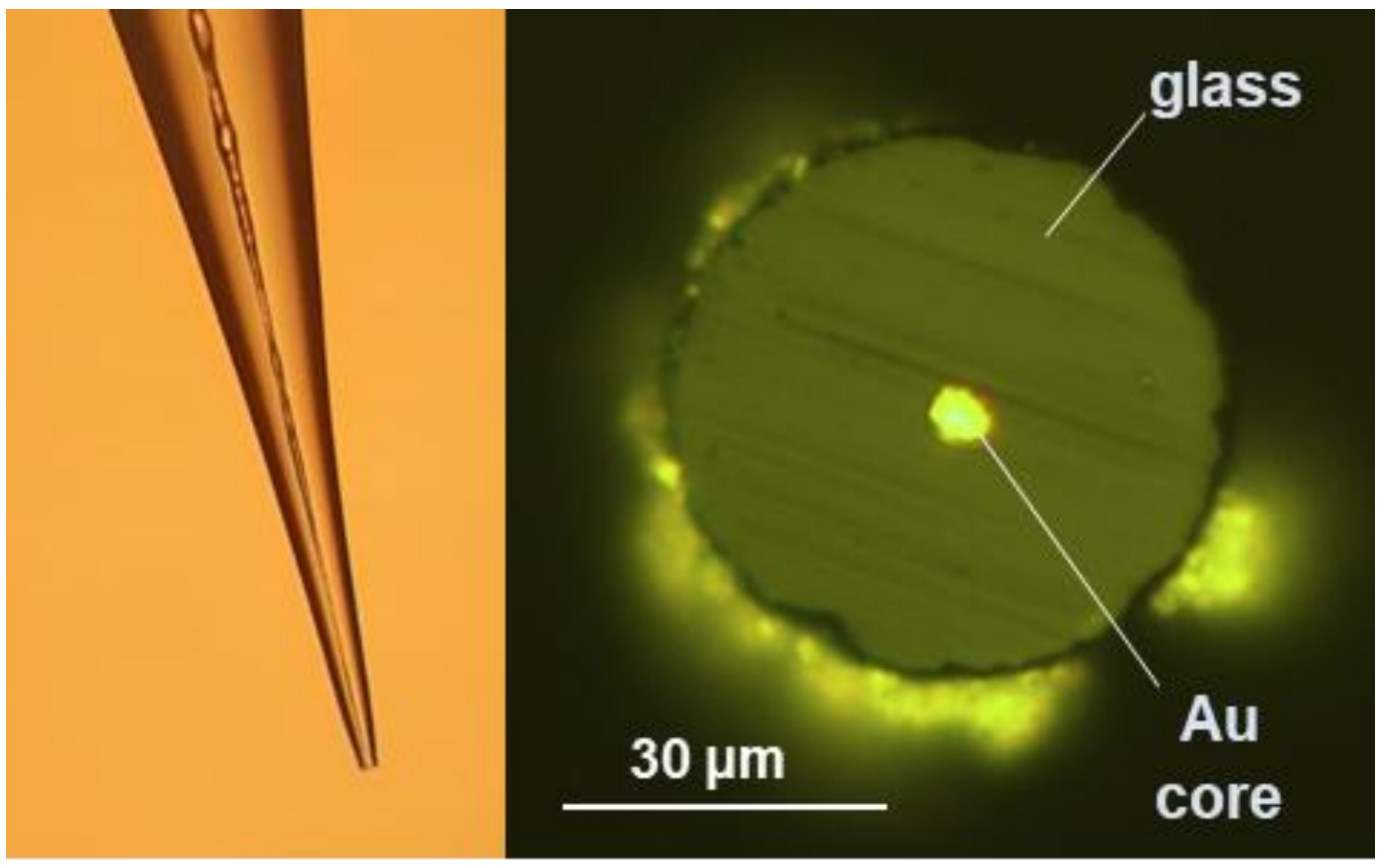

**Figure S4.** Optical microscopy images of the Au UME tip from the side and front views.

The UMEs discussed in this Note had the following structural characteristics:

| UME | Core radius (μm) | Glass radius (μm) | $R_G$ |
|---|---|---|---|
| Au | 1.5 | 24 | 16 |
| Pt | 7.5 | 52 | 7 |

Before oxygen-detection experiments, the tips were electrochemically conditioned in the measurement electrolyte by potential cycling (**Figure S5**).

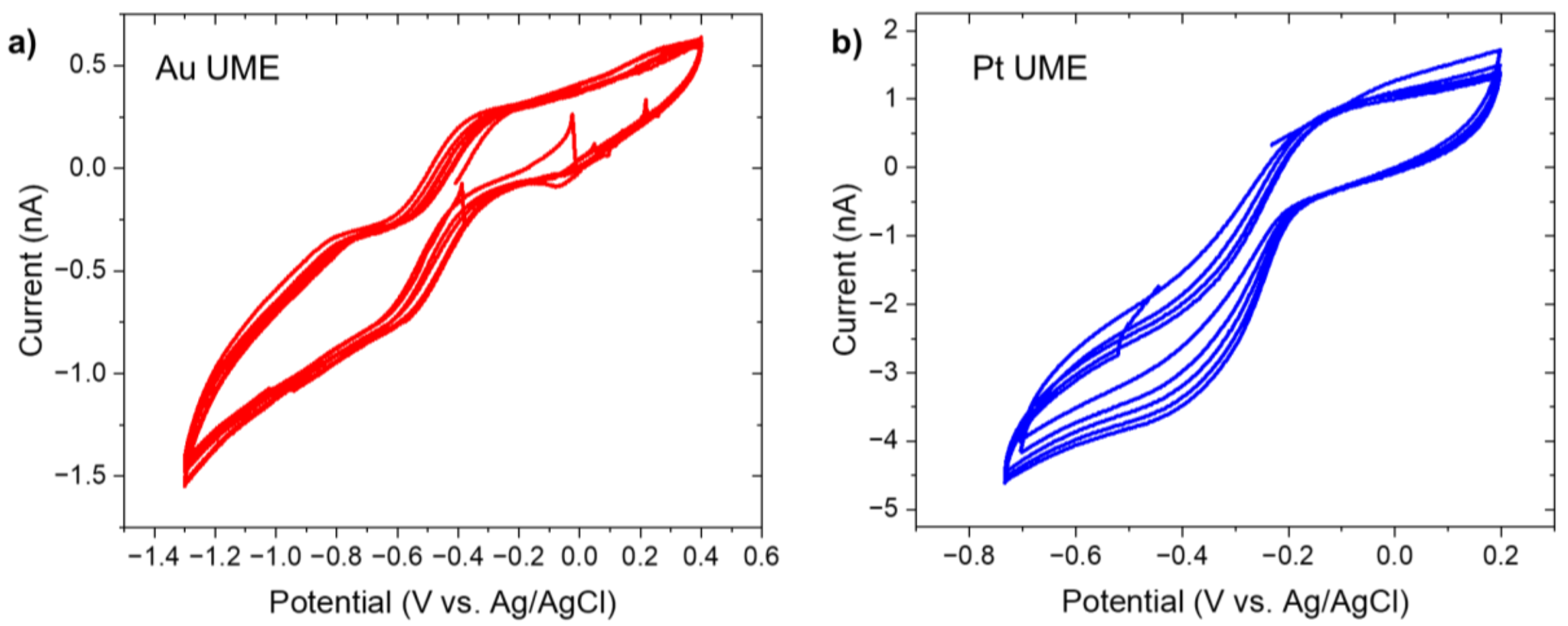

**Figure S5.** Cyclic voltammograms recorded in 0.1 M KOH during electrochemical conditioning of a) Au and b) Pt UMEs photo-SECM experiments.

### *Stability comparison of Pt and Au UMEs during oxygen detection*

The temporal stability of the oxygen-reduction current was evaluated under the bias conditions used for photo-SECM. After 15 min of repeated oxygen-detection measurements, the photo-induced tip response of Pt UME decreased 35 %, whereas the Au UME decreased only 0.3%.

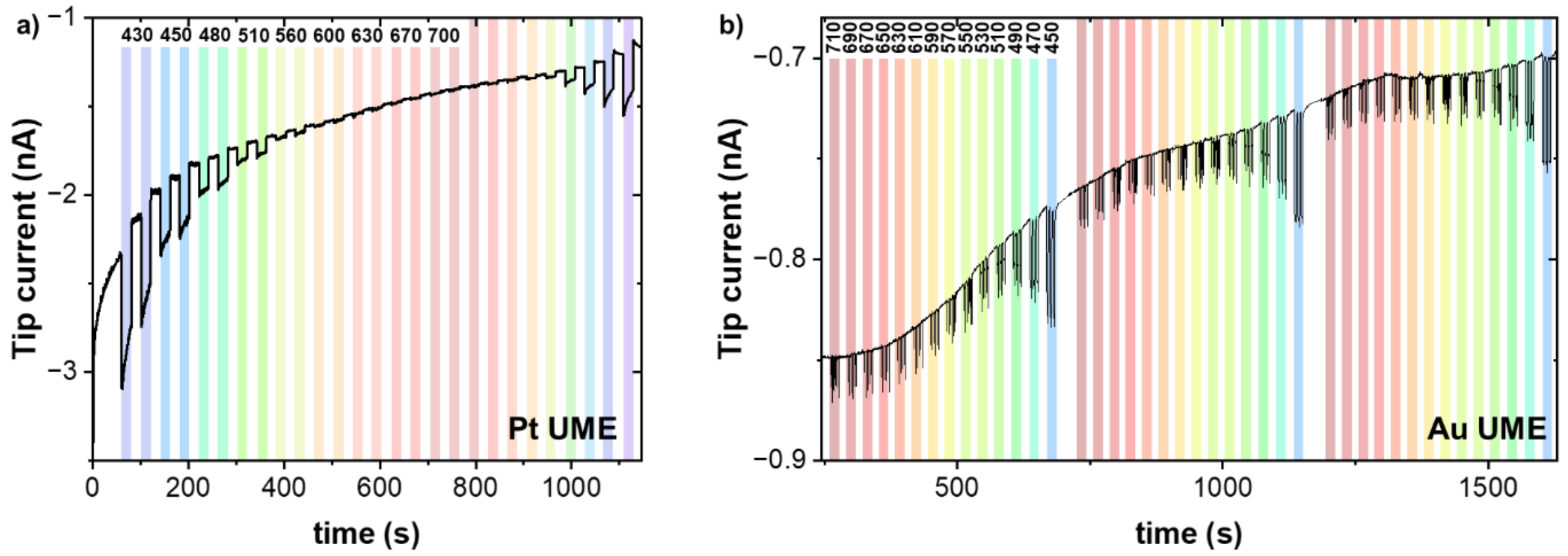

**Figure S6. Stability comparison of Pt and Au ultramicroelectrodes for oxygen detection. a)** Representative oxygen-reduction current recorded at a Pt UME during repeated photo-SECM measurements in 0.1 M KOH at a tip potential of -1 V vs Ag/AgCl. The Pt tip current decreases significantly with time, indicating instability under the measurement conditions. **b)** Representative oxygen-reduction current recorded at an Au UME. The Au tip maintains a more stable response over repeated measurements. Wavelengths (in nm) are represented by their respective colors as indicated in the first cycle.

Pt UMEs initially showed a measurable oxygen-reduction response at 0 V vs Ag/AgCl, but the difference in tip current between on and off cycles decreased significantly during repeated measurements. This current decay was observed under different tested conditions ($H_2SO_4$ 0.5M, $Na_2CO_3$ 0.1 M and KOH 0.1 M). The loss of response over time indicates that Pt tips were not sufficiently stable for comparison of oxygen evolution rates across different wavelengths or electrode modifications. Possible causes include surface oxidation/reduction, adsorption of electrolyte species or reaction intermediates, fouling by dissolved species, or changes in the Pt surface state during repeated oxygen-reduction measurements.

In contrast, Au UMEs exhibited a more stable ORR current over the same measurement period. The light-induced oxygen-detection signal remained comparatively constant during repeated illumination cycles, allowing more reliable comparison between measurements. Au tips were therefore selected for all quantitative photo-SECM measurements reported in the main text.

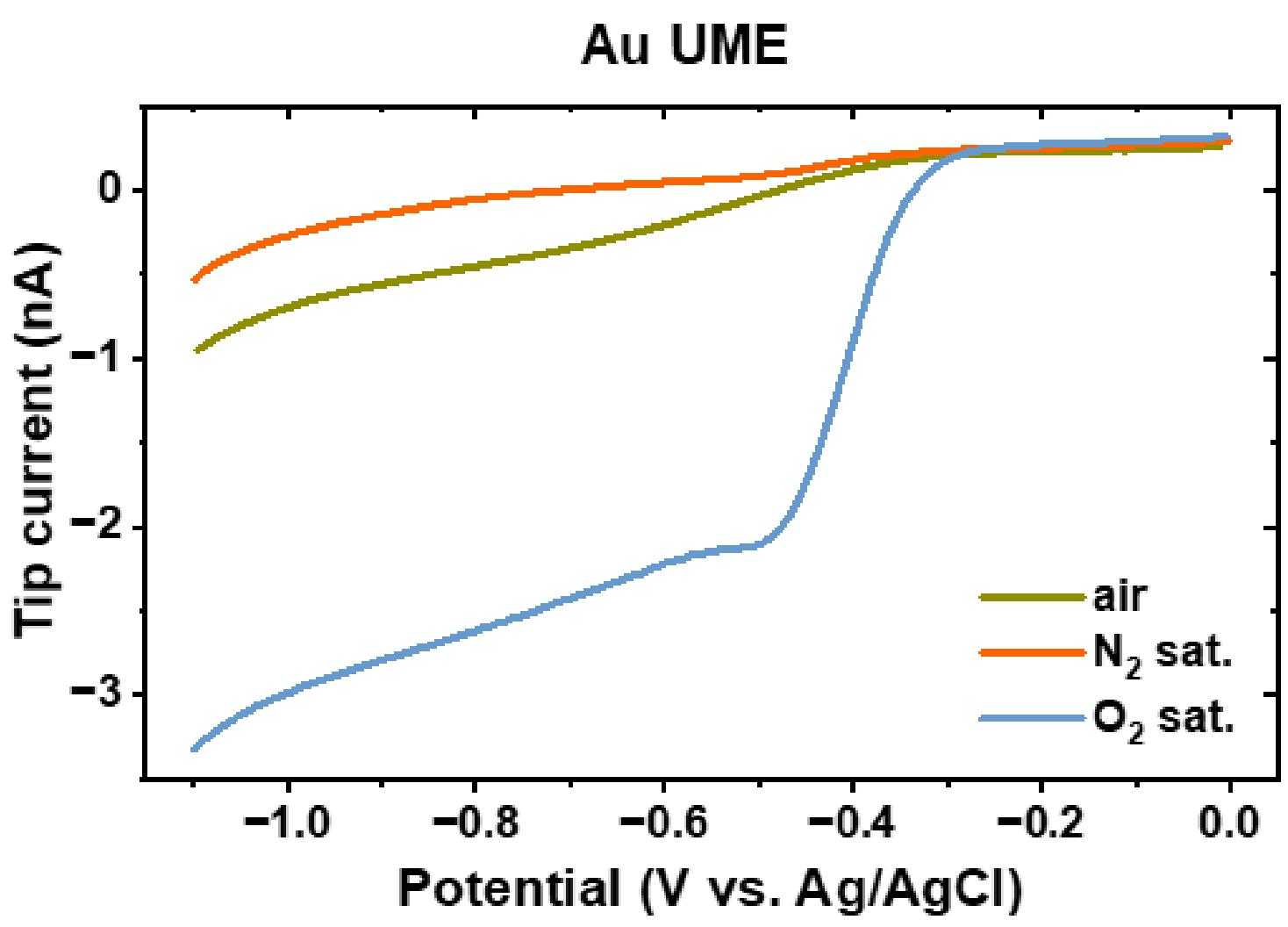

**Figure S7.** Cyclic voltammograms of the Au UME recorded in air-, $N_2$- and $O_2$-saturated 0.1 M KOH, showing the oxygen reduction response of the tip.

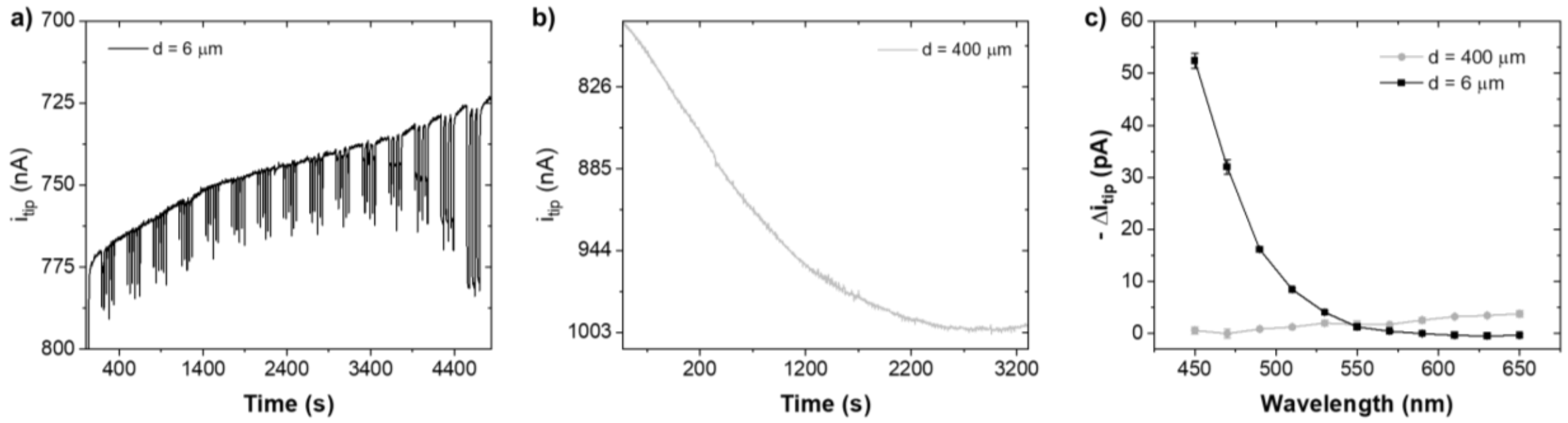

**Figure S8. Tip response under illumination.** To confirm that the measured tip response originates from $O_2$ generated at the illuminated photoanode rather than from a direct interaction between the UME and the incident light, a distance-dependent control experiment was performed. **a)** The tip was first positioned 6 µm above a $TiO_2$ substrate and the current response was recorded under wavelength-dependent chopped illumination (3 cycles at 710 to 450 nm every 20 nm, 100 µW). Under these conditions, the tip current showed a clear wavelength-dependent response, consistent with local detection of $O_2$ produced at the illuminated substrate. **b)** The same measurement was then repeated with the tip positioned far from the substrate (~400 µm). At this larger tip–substrate distance, no significant illumination-dependent tip current was observed. This confirms that the tip signal measured near the substrate arises from locally generated $O_2$ rather than direct photoresponse, heating, or optical perturbation of the tip. **c)** Comparison of -$\Delta i_{tip}$ in both scenarios.

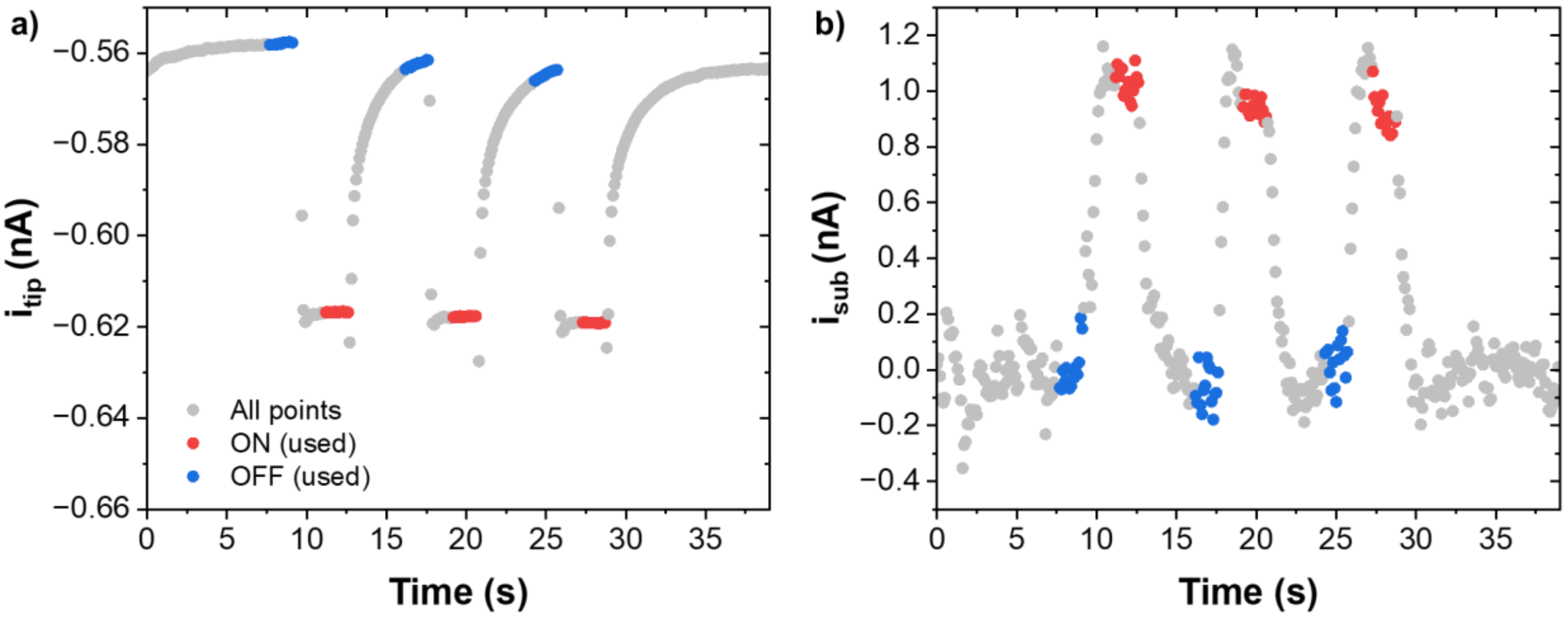

**Supplementary Figure S9. Extraction of $\Delta i_{sub}$ and $\Delta i_{tip}$ from chopped illumination traces**. Example time traces showing how $\Delta i_{sub}$ and $\Delta i_{tip}$ were extracted from chopped-illumination measurements. For each illumination cycle, the last N data points of the light-off period and the last N data points of the subsequent light-on period were selected after the transient response had stabilized. The same time windows were used for both the substrate-current and tip-current traces. The selected points within each window were averaged to obtain $i_{off}$ and $i_{on}$ for each cycle. The light-induced current change for each cycle was then calculated as $\Delta i = i_{on} - i_{off}$, using the immediately preceding off period as the baseline. The final reported $\Delta i_{sub}$ and $\Delta i_{tip}$ values correspond to the average of the three on–off differences.

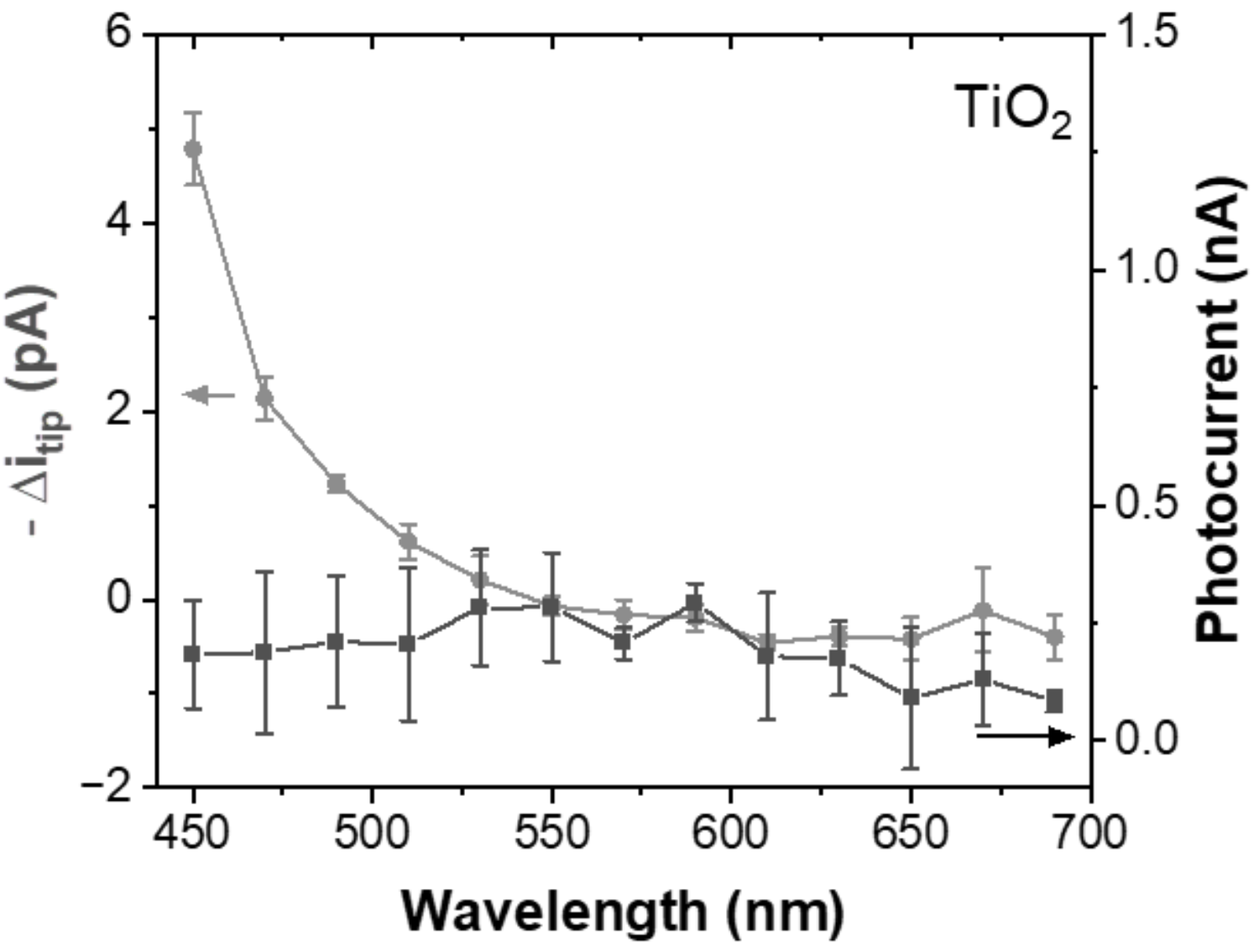

**Supplementary Figure S10.** Wavelength-dependent substrate photocurrent ($\Delta i_{sub}$) and corresponding $-\Delta i_{tip}$ values for bare $TiO_2$.

## Supplementary Note 2. Selection of suitable power range for oxygen detection

The incident-power range suitable for quantitative photo-SECM oxygen detection depended on the individual Au ultramicroelectrode used. Although the nominal diameter of the Au UMEs was approximately 4 μm, small differences in the exposed Au radius, glass sheath geometry, polishing history, and oxygen-reduction sensitivity resulted in different absolute currents and collection responses between tips. Consequently, the illumination power was determined separately for each UME before wavelength-dependent measurements.

The operating power was selected using the following criteria: (i) reproducible tip-current changes over consecutive illumination cycles; (ii) establishment of a stable current during each illumination period; and (iii) recovery of the dark baseline after illumination. Signals that did not satisfy these criteria were considered affected by nonlinear oxygen transport, local accumulation, convection, bubble formation, or changes in the tip response and were not used for quantitative wavelength-dependent comparisons.

For the Au UME used to characterize the unmodified $TiO_2$/Au/NiFe photoanodes (*Tip A*), illumination at 1 μW produced stable and reproducible oxygen-collection currents. With this tip, measurements at 10 μW showed sharp transients, decay during illumination, and incomplete recovery of the dark baseline (**Figure S11**). Therefore, 1 μW was selected for the measurements performed with *Tip A*. The cysteine-modified photoanodes were measured in a separate experimental series using a different Au UME (*Tip B*). *Tip B* maintained reproducible illumination-dependent currents at 10 μW. Therefore, 10 μW was used for the direct comparison of the L-cys- and DL-cys-modified samples.

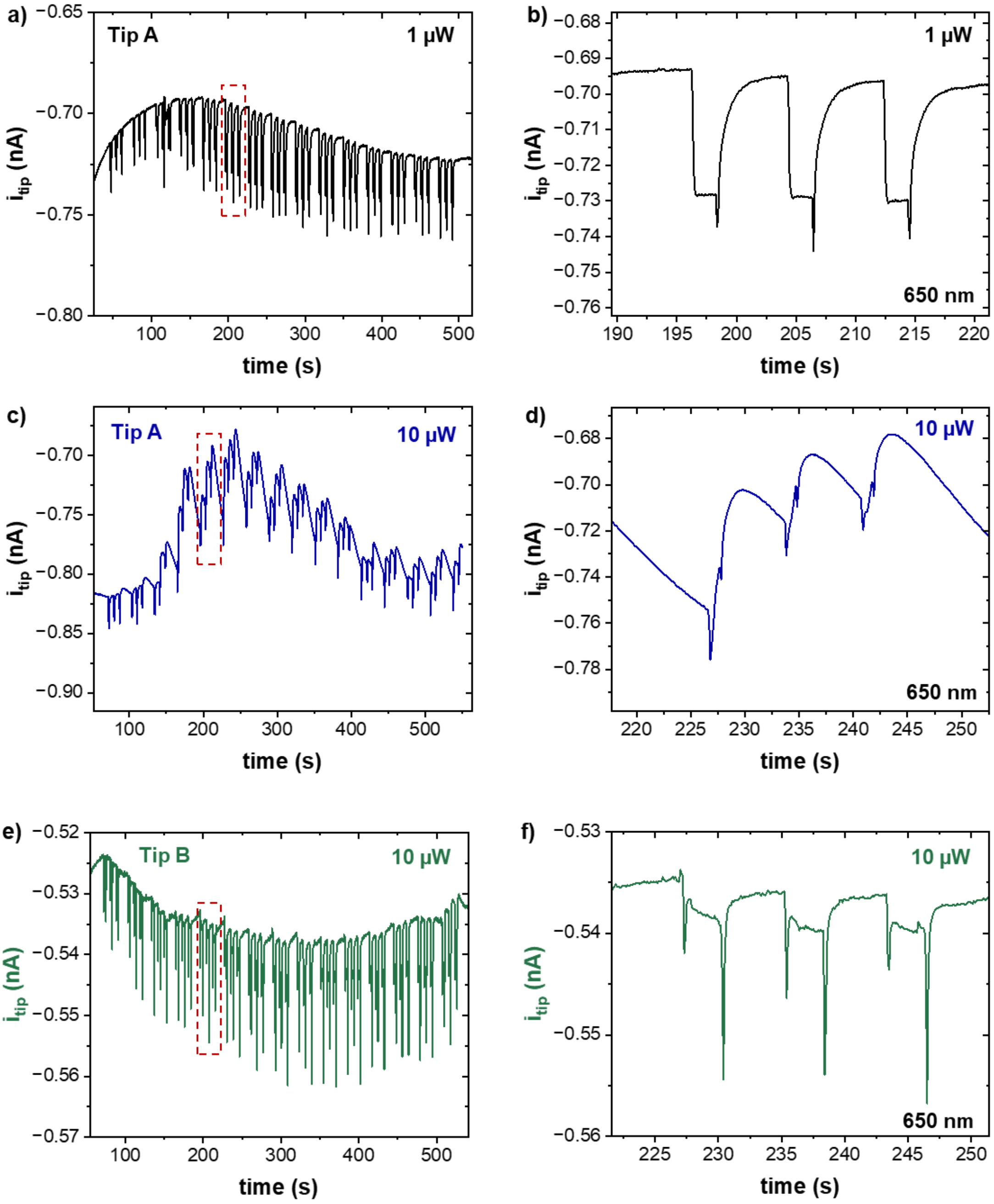

**Figure S11. Power dependence of the photo-SECM $O_2$ collection response in SG/TC mode for.** a–d) Tip A current measured using excitation powers of 1 µW (a,b) and 10 µW (c,d), when the excitation wavelength was varied from 750 to 450 nm in 20 nm steps, with three illumination cycles at each wavelength. e,f) Tip B current measured using excitation power of 10 µW. Panels (b), (d) and (f) show magnified views of the regions marked in red, corresponding to the response at 650 nm.

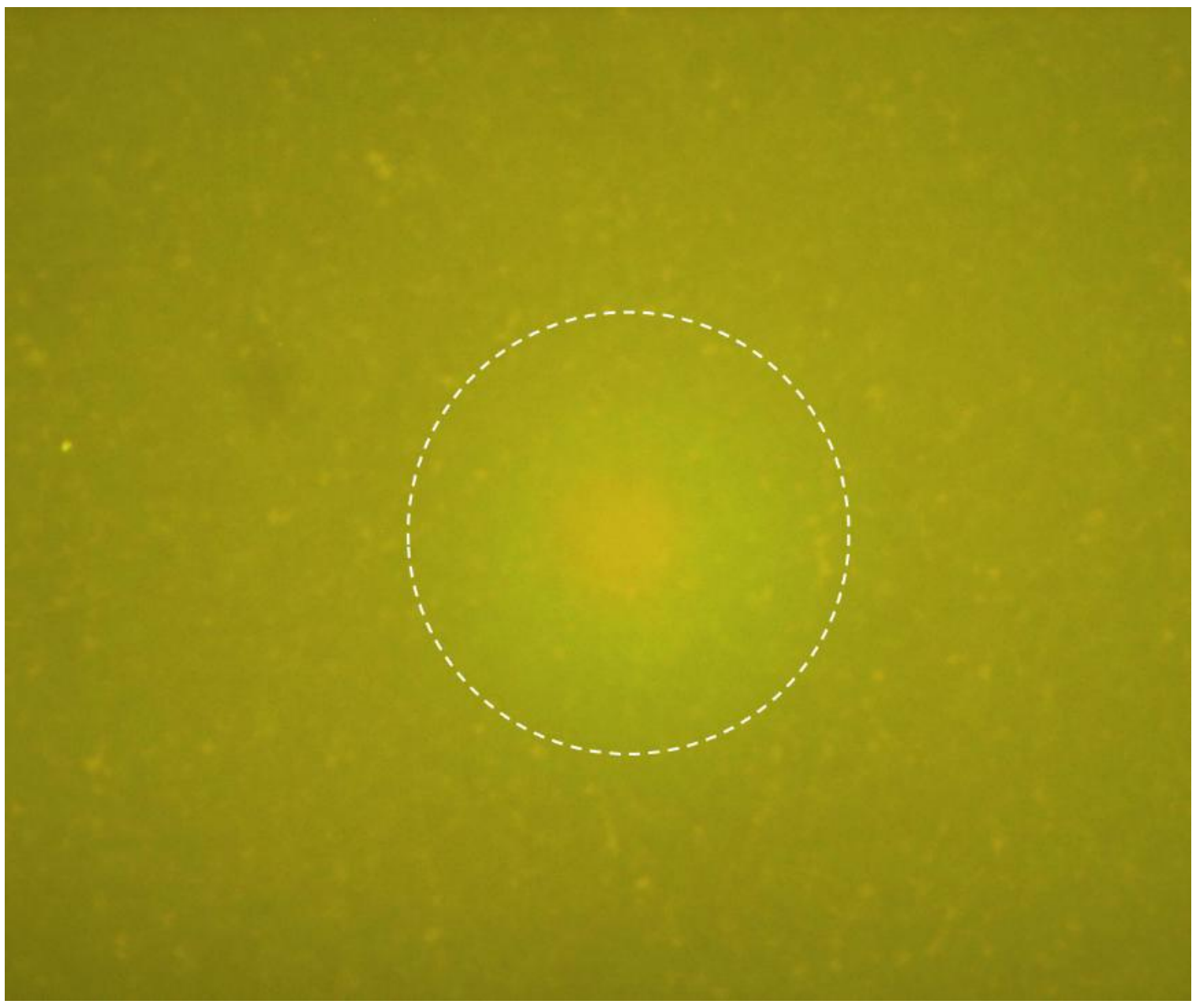

**Figure S12.** Optical microscopy image showing a color change in the illuminated region of the sample after laser excitation.

**Supplementary Note 3. Internal Quantum Efficiency (IQE) calculations.**

IQE values were calculated from the photocurrent response and the optical absorptance of the photoanodes. Although the photocurrent recorded during photo-SECM provides a direct correlation between substrate current and tip-detected $O_2$ evolution, additional photocurrent measurements were performed at higher incident powers to improve the signal-to-noise ratio for IQE determination (**Figure S11**). This was necessary because the illuminated region was limited to a ca. 30 µm-diameter spot, while the electrically contacted photoanode area was close to 1 $cm^2$, resulting in very small absolute photocurrents under the low-power photo-SECM conditions. The wavelength-dependent photocurrent exhibited the same spectral trend at the studied incident powers, and linear response was observed with respect to the power, suggesting that there are no significant photothermal contributions under the conditions used for IQE determination (**Figure S12**).

The IQE was then obtained from:

$$IQE = \frac{collected\ charge\ carriers}{absorbed\ photons} = \frac{\frac{|\Delta i_{sub}|}{q}}{\frac{P_{inc}\ A}{hc/\lambda}} = \frac{1240\ |\Delta i_{sub}|}{\lambda\ P_{inc}\ A}$$

where $\Delta i_{sub}$ is the substrate photocurrent, $q$ is the elementary charge, $h$ is Planck's constant, $c$ is the speed of light, $P_{inc}$ is the incident optical power at the sample at each wavelength ($\lambda$), and $A$ is the locally measured absorptance.

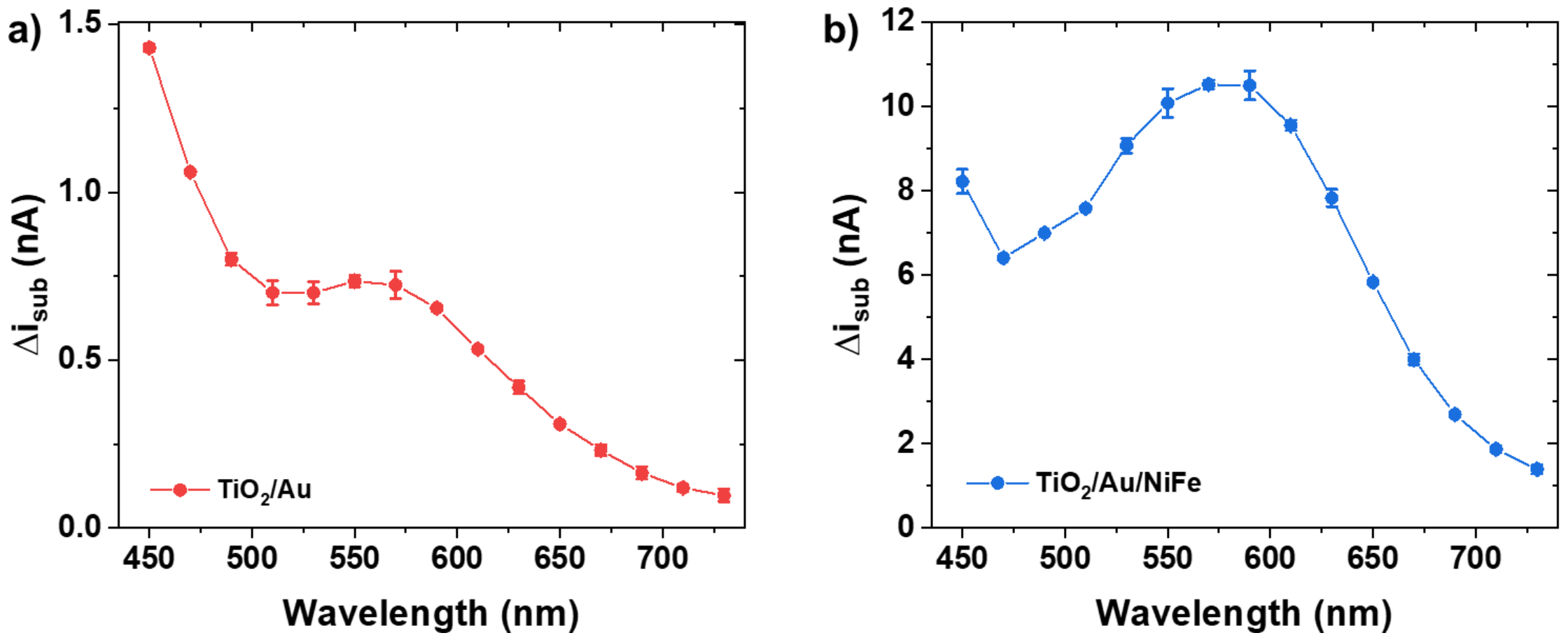

**Figure S13.** Photocurrent values used for IQE analysis.

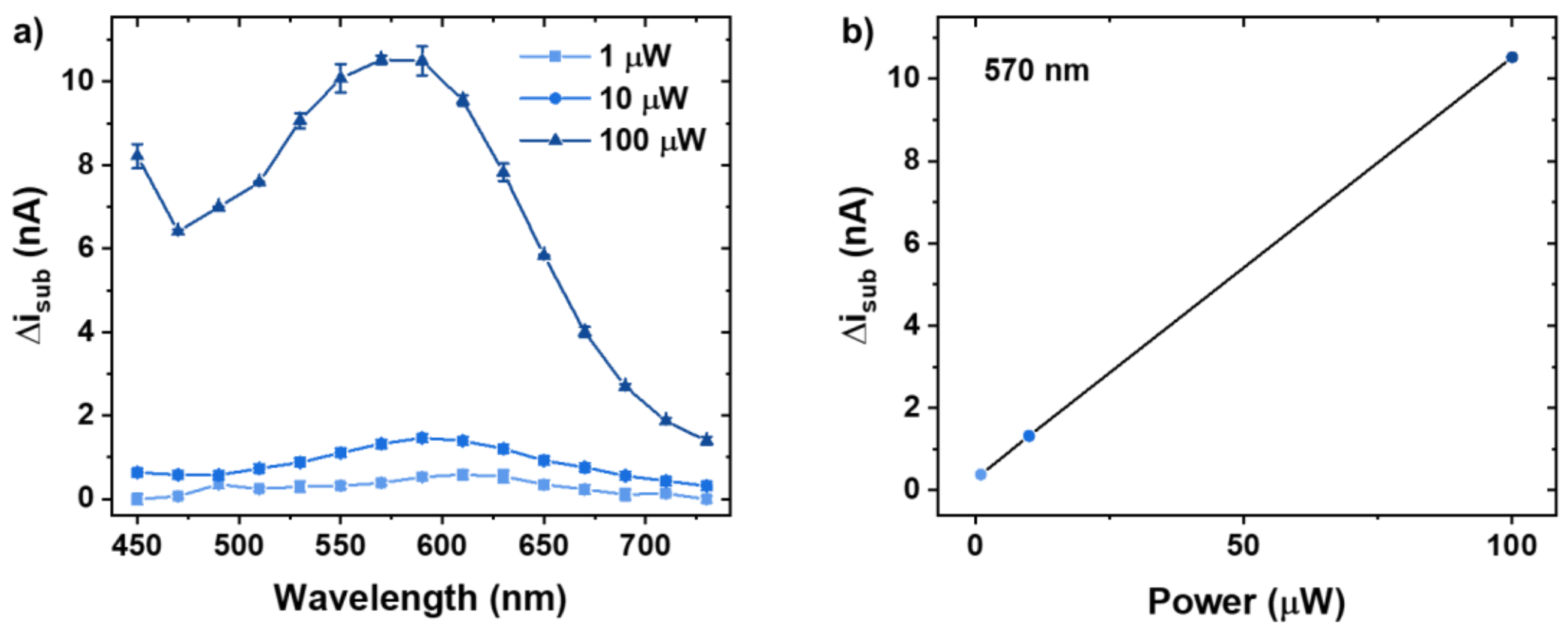

**Figure S14.** a) Photocurrent of $TiO_2$/Au/NiFe at increasing power. b) Linear increase of the photocurrent at 570 nm. Fit gives $R^2$ > 0.999.